\documentclass[%
  reprint,
 amsmath,amssymb,
 aps,
]{revtex4-2}
\usepackage[pdftex]{graphicx}
\usepackage{graphicx}
\usepackage{xcolor}
\usepackage{ulem}
\graphicspath{{./figs/}}
\usepackage{dcolumn}
\usepackage{bm}
\usepackage{multirow}
\usepackage{here}
\usepackage{empheq}

\usepackage[bookmarks = true, pdfpagemode = None, pdfstartview = FitH, colorlinks = true, urlcolor = blue]{hyperref}
\usepackage[braket, qm]{qcircuit}
\usepackage{amsmath}  
\usepackage{listings}

\newcommand\Erase{\bgroup\markoverwith{\textcolor{red}{\rule[.5ex]{2pt}{0.4pt}}}\ULon}
\newcommand\Eraseadd{\bgroup\markoverwith{\textcolor{red}{\rule[.5ex]{2pt}{0.4pt}}}\ULon}

\begin{document}

\preprint{APS/123-QED}

\title{Fermionic quantum approximate optimization algorithm}

\author{Takuya Yoshioka$^1$, Keita Sasada$^1$, Yuichiro Nakano$^2$, and Keisuke Fujii$^{2,3,4}$}
\affiliation{$^1$Strategic Technology Center, TIS Inc., 2-2-1 Toyosu, Koto-ku, Tokyo 135-0061, Japan}%
\affiliation{$^2$Graduate School of Engineering Science, Osaka University, 1-3 Machikaneyama, Toyonaka, Osaka 560-8531, Japan.
}%
\affiliation{$^3$Center for Quantum Information and Quantum Biology, Institute for Open and Transdisciplinary Research Initiatives, Osaka Univrsity, 1-2 Machikaneyama, Toyonaka, Osaka 560-0043, Japan.}
\affiliation{$^4$Center for Quantum Computing, RIKEN, 2-1 Hirosawa, Wako, Saitama 351-0198, Japan.}


\date{\today}

\begin{abstract}
  
Quantum computers are expected to accelerate solving combinatorial optimization 
problems, including algorithms such as Grover adaptive search and quantum 
approximate optimization algorithm (QAOA). However, many combinatorial optimization 
problems involve constraints which, when imposed as soft constraints in the cost 
function, can negatively impact the performance of the optimization algorithm. In this 
paper, we propose fermionic quantum approximate optimization algorithm (FQAOA) for 
solving combinatorial optimization problems with constraints. Specifically FQAOA tackle 
the constrains issue by using fermion particle number preservation to intrinsically 
impose them throughout QAOA. We provide a systematic guideline for designing the 
driver Hamiltonian for a given problem Hamiltonian with constraints. The initial state 
can be chosen to be a superposition of states satisfying the constraint and the ground 
state of the driver Hamiltonian. 
This property is important since FQAOA reduced to quantum adiabatic computation in 
the large limit of circuit depth $p$ and improved performance, even for shallow circuits with 
optimizing the parameters starting from the fixed-angle determined by Trotterized 
quantum adiabatic evolution. We perform an extensive numerical simulation and 
demonstrates that proposed FQAOA provides substantial performance advantage 
against existing approaches in portfolio optimization problems. Furthermore, the 
Hamiltonian design guideline is useful not only for QAOA, but also Grover adaptive 
search and quantum phase estimation to solve combinatorial optimization problems 
with constraints. Since software tools for fermionic systems have been developed in 
quantum computational chemistry both for noisy intermediate-scale quantum 
computers and fault-tolerant quantum computers, FQAOA allows us to apply these 
tools for constrained combinatorial optimization problems.
  
\end{abstract}

\maketitle



\section{Introduction}
Quantum optimization algorithms have attracted attention because of the potential for quantum computation to establish advantages in practical problems.
The targets are combinatorial optimization problems in industry, such as financial optimization \cite{Rosenberg, Hodson}, logistics and distribution optimization \cite{Streif, Stuart}, and energy optimization \cite{Ajagekar}.
In these practical problems, it is necessary to find a combination that gives the minimization cost while satisfying the constraints.

A standard quantum approach to solving these optimization problems are quantum annealing based on the adiabatic theorem \cite{Kadowaki, Farhi1, Aharonov}.
This approach slowly transforms a system Hamiltonian from a driver Hamiltonian $\hat{\cal H}_d$ to a cost function-based problem Hamiltonian $\hat{\cal H}_p$,
  which leads to optimal solutions of the cost function from the ground state (g.s.) of $\hat{\cal H}_d$.
However, the problems that can be solved with quantum annealing are limited to quadratic unconstrained binary optimization (QUBO) problems,
whereas in practical industrial problems some constraints are imposed on general variables that are not restricted to binary values.  
Thus, in quantum annealing,
a penalty function of quadratic form must be incorporated into the $\hat{\cal H}_p$ as a soft constraint.
Since this constraint is treated as an approximation, it may occasionally encounter states that are out of the solution space,
making the computation inaccurate \cite{Niroula, Wang, Hodson}.
  Furthermore, for problems with large realistic sizes and complex interactions, the first excitation gap in the adiabatic time evolution becomes smaller and the adiabatic dynamics is time consuming \cite{Young}.

Next, we will introduce the quantum approximate optimization algorithm (QAOA) \cite{Farhi2},
  a variational algorithm for solving combinatorial optimization problems utilizing the controllability of a universal quantum computer.
In principle, there are no hardware connectivity limitations, and quantum gates can handle higher-order interactions as well as second-order \cite{Gilliam}, which allows efficient encoding of general variables \cite{Sano}.
This method covers the quantum adiabatic algorithm (QAA) by taking a large circuit depth $p$ \cite{Farhi1, Farhi2}.
Owing to this property, the variational parameters can be presumed to some extent,
  which may be an advantage even in shallow circuits \cite{Cao, Hastings, Wurtz}.
  However, the same problems as in quantum annealing exist here for constrained optimization problems.
  
To solve these problems, Hadfield $et$ $al$. proposed a hard constraint approach: a quantum alternating operator ansatz \cite{Hadfield1, Hadfield2},
which enforces the generated quantum states into the feasible subspace.
In particular, $XY$-QAOA using a mixer with an $XY$ Hamiltonian \cite{Lieb} has been applied to graph-coloring problems \cite{Wang},
portfolio optimization problems \cite{Hodson}, extractive summarization problems \cite{Niroula}, etc.
In these problems, constraints are imposed on the bit strings that keep the hamming weight constant,
and the initial states are based on Dicke states \cite{Dicke} that satisfy the constraints.

According to these previous studies, the algorithms
improve the computational accuracy by restricting the combinatorial search space, however, they have the following common problems.
  Due to generality of the mixer, there is no systematic description of an ansatz incorporating the hard constraints,
  and the relations between the initial states and the driver Hamiltonian are
  not explored enough.
  Therefore, the algorithm does not guarantee that it will result in QAA in the limit of large $p$.
  In addition, although the number of variational parameters remains the same as in conventional methods, no special effort has been made to handle the parameter optimization.
  Therefore, a systematic and efficient method of solving constrained optimization problems with guaranteed convergence is required.

In this paper, we developed a systematic framework called fermionic QAOA (FQAOA) for constrained optimization problems using fermionic formulation.
The FQAOA translates the problem Hamiltonian and the driver Hamiltonian
of a conventional spin system into the representation of fermion systems,
and the equality constraint is naturally incorporated as a constant number of particles condition, and hence no penalty function is needed.
We also show that the g.s. of the driver Hamiltonian can be prepared as an initial state, thus FQAOA will be QAA in the large $p$ limit,
allowing us to execute the time-discretized QAA.
  The parameters used here can then be set as initial parameters for
  variational quantum algorithms and efficiently computed using gradient descent methods, including parameter shifting methods \cite{Li, Mitarai}.
This FQAOA framework is reduced to QAA in the limit of large $p$.

As a representative example, we will take a portfolio optimization problem.
This is a second-order, equality-constrained, three-variable problem.
Numerical simulations demonstrate that the proposed FQAOA provides a significant performance improvement.
In particular, the computational accuracy at $p=1$ of FQAOA with fixed-angles determined by Trotterized QAA outperforms
the results of the $XY$-QAOA using parameter optimization at $p=4$ in the previous study \cite{Hodson}
by about half of gate operations.
In addition, the FQAOA with parameter optimization at $p=4$ improves the probability of achieving low-energy states by a factor of 40
compared to the results of $XY$-QAOA at $p=4$ in Ref. \cite{Hodson}.
Our algorithm enables us to simulate a wide variety of constraint optimization problems with high accuracy
by using the platforms of quantum chemical computation both for noisy intermediate-scale quantum computers (NISQ) and fault-tolerant quantum computers (FTQC).

This paper is organized as follows. In Sec. II, we introduce the constrained polynomial optimization problems.
  We formulate the FQAOA for these problems in Sec. III and introduce the quantization of the problems
by the fermionic representation.
In Sec. IV, we present the general framework for portfolio optimization problems and
evaluate the numerical results of the FQAOA, 
where we compare our results with the computational results of other QAOAs, including the previous study.
A summary is provided in Sec. V.\\

\section{Constrained Optimization Problems}
The cost function $E(\bm z)$ for polynominal optimization problems with $N$ integer variables $z_l\in \{0, 1, 2, \cdots, I\}$ takes the following form:
\begin{eqnarray}    
  E({\bm z}) &=& \sum_{k=1}^K\sum_{\langle l_1, l_2,\cdots, l_k\rangle} \alpha_{l_1,l_2, \dots, l_k}z_{l_1}z_{l_2}\dots z_{l_k}, \nonumber\\
  {\rm s.t.}&& \sum_{l \in V_j} z_l= M_j\text{\qquad for $\forall$ $j$},\label{eq:constz}
\end{eqnarray}
where $\alpha_{l_1,l_2, \dots, l_k}$ represent $k$-body interactions and
$V_j$ is the $j$-th subset of vertex $l$. 
  $z_l$ can be embedded in binary $x_{l,d}\in\{0,1\}$ as \cite{Tamura, Rosenberg}
\begin{equation}
  z_l=\sum_{d=1}^D f_dx_{l,d},
\end{equation}
where encoding function $f_d$ is shown in Table \ref{tble:intenc}.
In this paper, we do not employ one-hot encoding since it introduces an extra constraint originating from the encoding shown in Table \ref{tble:intenc}.
For the remaining three encodings using $f_d$, the cost function is transformed into the following
\begin{eqnarray}      
E({\bm x})&=&\sum_{k=1}^K\sum_{\langle l_1, l_2,\cdots, l_k\rangle} \alpha_{l_1,l_2, \dots, l_k}
  \prod_{j=1}^k \left(\sum_{d_j=1}^Df_{d_j}x_{l_j,d_j}\right), \label{eq:cost}\\
{\rm s.t.}&&\sum_{l\in V_j}\sum_{d=1}^Df_dx_{l,d} = M_j\text{\qquad for $\forall$ $j$}.\label{eq:const}
\end{eqnarray}
Our goal is to find a bit string ${\bm x}^* = \underset{\bm x}{\arg \min} E({\bm x}) $
under the constraints in Eq. (\ref{eq:const}).

\begin{table}
  \caption{\label{tble:intenc}
Encoding function $f_d$ in $z_l=\sum_{d=1}^D f_dx_{l,d}$ \cite{Tamura, Rosenberg} and
the required number of bits $D$ to represent the largest integer $I$ of $z_l$, where $\lceil x \rceil$ is the ceiling function of $x$.
  }
\begin{ruledtabular}          
  \begin{tabular}{ c c c  c c}
Encoding & $f_d$ & $D$ & Requirement\\
\hline
Binary     & $f_d=2^{d-1}$  & $\lceil \log_2 (I+1) \rceil$ &\\
Unary      & $f_d=1$        & $I$  & \\
Sequential & $f_d=d$        & $\lceil[\sqrt{(1+8I)}-1]/2\rceil$&\\
One-hot    & $f_d=d$        & $I+1$ & $\sum_{d=1}^Dx_{l,d} = 1$ 
\end{tabular}
\end{ruledtabular}        
\end{table}

\section{Framework of Fermionic QAOA}

\begin{figure*}
  \includegraphics[width=17cm]{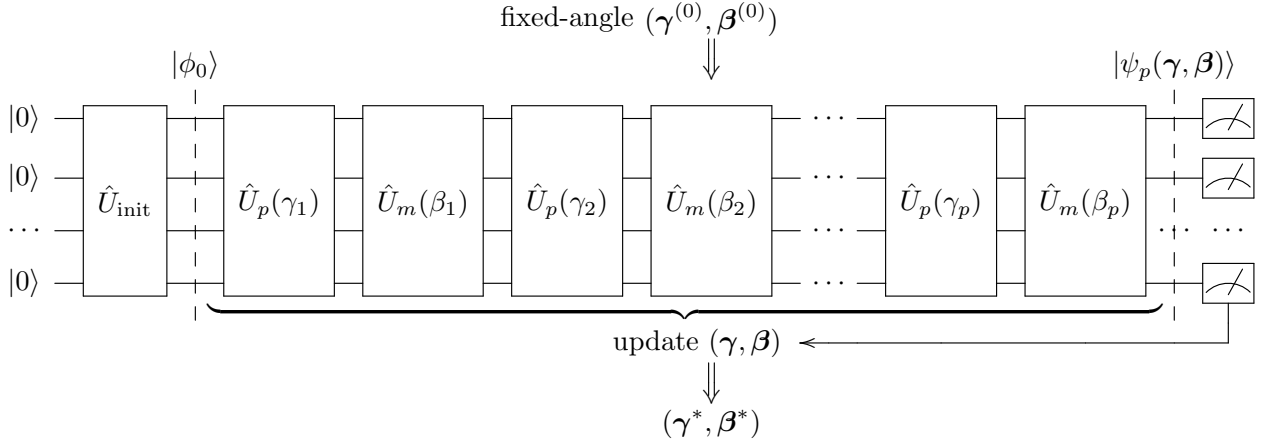}
  \caption{\label{fig:QAOA} Framework of FQAOA,
      where (${\bm \gamma}^{(0)}, {\bm \beta}^{(0)}$) are the parameters in the time-discretized QAA
      and (${\bm \gamma}^*, {\bm \beta}^*$) are the parameters resulting from the parameter update,
      and $\hat{U}_p(\gamma)=\exp(-i\gamma\hat{\cal H}_p)$, $\hat{U}_m(\beta)=\exp(-i\beta\hat{\cal H}_d)$, and
      $\hat{U}_{\rm init}$ is the unitary operator for preparing the ground state of the driver Hamiltonian $\hat{H}_d$. 
}  
\end{figure*}

We construct a framework of FQAOA shown in Fig. \ref{fig:QAOA},
which is a hybrid quantum-classical algorithm for solving constraint optimization problems
expressed in fermion form within the unary encoding.
The cost function and constraints in Eqs. (\ref{eq:cost}) and (\ref{eq:const}) are mapped to
the problem Hamiltonian $\hat{\cal H}_p$ 
with a fixed number of localized fermions.
A parametrized quantum circuit with $({\bm \gamma}, {\bm \beta})$ is used to compute the expectation value $E_p({\bm \gamma}, {\bm \beta})$ of $\hat{\cal H}_p$, which is minimized by the outer parameter update loop.
An FQAOA ansatz $\ket{\phi_p({\bm \gamma}, {\bm \beta})}$ consists of an initial state preparation unitary $\hat{U}_{\rm init}$ and a phase rotation unitary $\hat{U}_p(\gamma)$ and a mixing unitary $\hat{U}_m(\beta)$, 
\begin{eqnarray}
  \hat{U}_p(\gamma) &=& \exp(-i\gamma\hat{\cal H}_p),\label{eq:Up}\\
  \hat{U}_m(\beta) &=& \exp( -i\beta\hat{\cal H}_d), \label{eq:Ud}
\end{eqnarray}
where $\hat{\cal H}_d$ is the driver Hamiltonian describing the non-local fermions.
By carefully designing $\hat{\cal H}_d$ and $\hat{U}_{\rm init}$, the constraints are satisfied at any steps without additional penalty function.
As will be explained below, the framework is constructed so it comes down to QAA in the large $p$ limit.
The components of the FQAOA ansatz are described in detail in the following subsections.

\subsection{Mapping to fermionic formulation}
\begin{figure}[htb]
  \includegraphics[width=5cm]{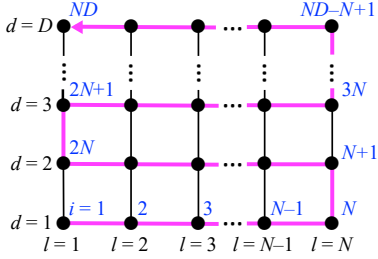}
  \caption{\label{fig:JW}
       The orderings of operators in the Jordan-Wigner (JW) encoding \cite{Jordan} used in this paper,
       so-called snake type JW ordering, 
       where sequential numbers $i=1, 2,...,ND$ shown in blue are assigned in the order according to Eqs. (\ref{eq:dl}) and (\ref{eq:idl}).
  }
\end{figure}

In this paper, the binary $x_{l,d}$ are mapped to the number operators of fermions on the ($l, d$) site as
\begin{eqnarray}
  x_{l,d}\longmapsto\hat{n}_{l,d}= \hat{c}_{l,d}^\dagger\hat{c}_{l,d},
\end{eqnarray}
where $\hat{c}^{\dagger}_{l,d}$($\hat{c}_{l,d}$) is the creation (annihilation) operator on ($l,d$) site,
which satisfies the anti-commutation relations:
  \begin{equation}
    \begin{split}
  \hat{c}_{l,d}\hat{c}^\dagger_{l',d'}+\hat{c}^\dagger_{l',d'}\hat{c}_{l,d}&=\delta_{l,l'}\delta_{d,d'},\\
  \hat{c}_{l,d}\hat{c}_{l',d'}+\hat{c}_{l',d'}\hat{c}_{l,d}&=\hat{c}^\dagger_{l,d}\hat{c}^\dagger_{l',d'}+\hat{c}^\dagger_{l',d'}\hat{c}^\dagger_{l,d}=0.\label{eq:antcom}
  \end{split}
\end{equation}
The computational basis corresponding to the bit string $\bm x\in\{0,1\}^{ND}$ can be written
\begin{eqnarray}
 \ket{\phi_{\bm{x}}}&=&\left(\hat{c}_{ND}^{\dagger}\right)^{x_{ND}}\cdots
 \left(\hat{c}_{2}^{\dagger}\right)^{x_2}\left(\hat{c}_{1}^{\dagger}\right)^{x_1}\ket{{\rm vac}},\label{eq:bit}\\
 \hat{n}_i\ket{\phi_{\bm x}}&=&x_i\ket{\phi_{\bm x}} \text{\qquad for  $\forall$ $i$},
\end{eqnarray}
where the subscript $(l,d)$ is collectively denoted as $i$ and
$\ket{{\rm vac}}$ is a vacuum satisfying $\hat{c_i}\ket{{\rm vac}} = 0$.
Note that the operators have the anti-commutation relations in Eq. (\ref{eq:antcom}), 
so the sequence of operators in Eq. (\ref{eq:bit}) follows a fixed order.
In this paper, we apply the snake type Jordan-Wigner (JW) ordering shown in Fig. \ref{fig:JW},
  in which the relation between $i$ and $(l, d)$ can be explicitly denoted by
\begin{equation}
    \begin{split}
  d_i&=\left\lceil\frac{i}{N}\right\rceil,\\
  l_i&=\frac{N+1-(-1)^{d_i}(N-1)}{2}+(-1)^{d_i}(Nd_i-i),
  \end{split}\label{eq:dl}
\end{equation}
for $i=1,2,\cdots ND$ and conversely
\begin{equation}
  i_{l,d}=(-1)^{d-1}l+(d-1)N+\frac{1-(-1)^{d-1}}{2}(N+1).\label{eq:idl}
\end{equation}

\subsection{Problem Hamiltonian  with linear constraints}
The cost function with constraints in Eq. (\ref{eq:cost}) with (\ref{eq:const})
are mapped to the following eigenvalue problems:
\begin{eqnarray}    
  \hat{\cal H}_p\ket{\phi_{\bm x}} &=& E({\bm x})\ket{\phi_{\bm x}},\label{eq:Hpeigen}\\
  \hat{C}_j\ket{\phi_{\bm x}}&=&M_j\ket{\phi_{\bm x}} \text{\qquad  for $\forall$ $j$},\label{eq:qconst}
\end{eqnarray}
where $\hat{\cal H}_p$ and $\hat{C}_j$ is problem Hamiltonian and $j$-th constraint operator, respectively,
which can be explicitly expressed as follows:
\begin{eqnarray}    
  \hat{\cal H}_p &=& \sum_{k=1}^K\sum_{\langle l_1, l_2,\cdots, l_k\rangle} \alpha_{l_1,l_2, \dots, l_k}
  \prod_{j=1}^k \left(\sum_{d_j=1}^D\hat{n}_{l_j,d_j}\right), \label{eq:Hp}\\
  \hat{C}_j&=&\sum_{l\in V_j}\sum_{d=1}^D\hat{n}_{l,d},
\end{eqnarray}
where $\hat{\cal H}_p$ is a model with $k=1, 2 \dots, K$-body interactions of localized fermions.
Thus, the constrained optimization problems have been replaced by the
  eigenvalue problems of obtaining the g.s. $\ket{\phi_{{\bm x}^*}}$ and its energy $E_{\rm min}=E({\bm x}^*)=\min_{\bm x}E({\bm x})$ under the constraints of Eq. (\ref{eq:qconst}).

\subsection{Driver Hamiltonian and initial states}
\label{sec:Hd}
We address the design of the driver Hamiltonian $\hat{\cal H}_d$ and initial states $\ket{\phi_0}=\hat{U}_{\rm init}\ket{\rm vac}$.
First, we assume that $[\hat{\cal H}_d,\hat{\cal H}_p]\ne 0$ to introduce hybridization between different basis states $\ket{\phi_{\bm x}}$,
Thus $\hat{\cal H}_d$ represent non-local fermions represented by hopping terms,
while $\hat{\cal H}_p$ denote a localized ones.
Conditions that $\hat{\cal H}_d$ and $\ket{\phi_0}$ must satisfy are as follows\\
\begin{itemize}
\item Condition I \\
  The following conditions are imposed as necessary conditions to satisfy the constraints at
  any approximation level $p$
  all times during the time evolution process.
\begin{equation}
  [\hat{\cal H}_d, \hat{C}_j]=0 \text{\qquad for $\forall$ $j$}.\label{eq:condI}
\end{equation}
\item Condition II\\
A series of hopping terms allows transitions between any two localized states $\ket{\phi_{{\bm x}'}}$ and $\ket{\phi_{\bm x}}$ that satisfy the constraints
\begin{equation}
 |\bra{\phi_{{\bm x}'}}(\hat{\cal H}_d)^n\ket{\phi_{\bm x}}| >0. \label{eq:condII}
\end{equation}
\item Condition III\\
  The initial state $\ket{\phi_0}=\hat{U}_{\rm init}$ is the g.s. of $\hat{\cal H}_d$
  and at the same time satisfies the constraints
\begin{eqnarray}
  \hat{\cal H}_d\ket{\phi_0} &=& E_{0}\ket{\phi_0},\label{eq:Hpeigen}\\  
  \hat{C}_j\ket{\phi_0}&=&M_j\ket{\phi_0} \text{\qquad  for $\forall$ $j$},\label{eq:qconstphi0}
  \end{eqnarray}
  where $E_0$ is the g.s. energy of $\hat{\cal H}_d$.
\end{itemize}
By using the designed $\hat{\cal H}_d$ and its g.s. as the initial state $\ket{\phi_0}$,
  the constrained optimization problems can be solved by performing adiabatic time evolution from the non-local initial state to the desired localized state.
  It is expected that FQAOA in this setting will allow more efficient solving of constrained optimization problems with shallow circuits.

\subsection{Fermionic QAOA ansatz}
The FQAOA ansatz satisfying the hard constraint is written in the following
\begin{eqnarray}
  \ket{\psi_p({\bm \gamma},{\bm \beta})}&=&\left[\prod_{j=1}^{p}\hat{U}_m(\beta_j)\hat{U}_p(\gamma_j)\right]\hat{U}_{\rm init}\ket{\rm vac}.
  \label{eq:Ansqaoa}
\end{eqnarray}
  The $\hat{U}_m(\gamma)$ is the mixer that changes the fermion configurations,
  the generator of which is the driver Hamiltonian $\hat{\cal H}_d$ satisfying condition I and II in Sec. \ref{sec:Hd}.
The initial state $\ket{\phi_0}=\hat{U}_{\rm init}\ket{\rm vac}$ satisfies condition III in Sec. \ref{sec:Hd}.
For efficient quantum computation, it is important that
$\hat{U}_{\rm init}$ can be implemented in polynomial time.
  We will see this is actually the case for a linear order in terms of the number of qubits
  in the next section with a concrete example of the portfolio optimization problem.

The FQAOA contains the QAA \cite{Farhi1} in the limit of large $p$.
Taking the time-dependent Hamiltonian as $\hat{\cal H}(t)=(1-t/T)\hat{\cal H}_d+(t/T)\hat{\cal H}_p$,
  discretizing time $t$ by $t_j=(2j-1)T/2p$ ($j=1, 2, \dots, p$) with the execution time $T=p\Delta t$, the parameters assigned are as follows:
\begin{equation}
  \begin{split}
  \gamma_j^{(0)} &= \frac{2j-1}{2p}\Delta t, \\  
  \beta_j^{(0)}  &= \left(1-\frac{2j-1}{2p}\right)\Delta t,\label{eq:gamma0beta0}
  \end{split}
\end{equation}
the derivation of which by QAA is shown in Appendix \ref{sec:qaa}.
As there, the key condition for performing the above QAA is that
the initial state must be the g.s. of $\hat{\cal H}_d$ under the constraint,
  which is satisfied by condition III.
Previous studies on constrained optimization problems \cite{Niroula, Wang} often use the Dicke state as the initial state, and some of them do not satisfy this condition.

\subsection{Parameter optimization for fermionic QAOA}
  The objective of FQAOA is to obtain a bitstring that gives the lowest energy.
  To this end, the energy expectation value $E_p({\bm \gamma},{\bm \beta})$ needs to be computed,
  which is obtained as the output of the FQAOA ansatz in Eq. (\ref{eq:Ansqaoa}) as
\begin{eqnarray}
    E_p({\bm \gamma},{\bm \beta})&=&\bra{\psi_p({\bm \gamma},{\bm \beta})}{\cal H}_p\ket{\psi_p({\bm \gamma},{\bm \beta})}.
  \label{eq:Eqaoa}
\end{eqnarray}
The minimum energy at the approximation level $p$ is determined by the following parameter optimization \cite{Farhi2}:
\begin{equation}
E_p(\bm{\gamma}^*,\bm{\beta}^*)=   \min_{\bm{\gamma},\bm{\beta}} E_p(\bm{\gamma},\bm{\beta}),\label{eq:optEp}
\end{equation}
where the parameter set $(\bm{\gamma}^*,\bm{\beta}^*)$ determines the approximate variational wave function of the g.s. as $\ket{\psi_p(\bm{\gamma}^*,\bm{\beta}^*)}$.
The wave function $\ket{\psi_p(\bm{\gamma},\bm{\beta})}$ determines
  a probability $P_{\bm x}(\bm{\gamma},\bm{\beta})$ that a bit string $\bm x$ is observed as
\begin{equation}
   P_{\bm x}({\bm \gamma},{\bm \beta})=\ket{\bra{\phi_{\bm x}}\psi_p({\bm \gamma},{\bm \beta})}|^2.\label{eq:Px} 
\end{equation}

In this paper, to avoid the difficulty of parameter optimization, the parameters in Eq. (\ref{eq:gamma0beta0}) are used as initial values for the parameter optimization.
The above mentioned calculation procedure is summarized in Fig.\ref{fig:QAOA}.
\\

\section{Portfolio Optimization Problem}
As an example of the application of FQAOA, we take a portfolio optimization problem
and evaluate its performance by comparing with a previous study \cite{Hodson}.
The cost function for the optimization problem
under the constraint of the total number of stock holdings being $M$ is defined by \cite{Markowitz, Rosenberg, Hodson}
\begin{eqnarray}
	E({\bm w}) &=& \frac{\lambda}{M^2} \sum_{l, l'=1}^N \sigma_{l,l'} w_lw_{l'} - \frac{ 1 - \lambda}{M} \sum_{l=1}^N \mu_l w_l,
        \label{eq:Marko}\\
        {\rm s.t.}&&\sum_{l=1}^N w_l = M, \label{eq:CpenM}
\end{eqnarray}
where the subscripts $l$ and $l'$ are the indices of the stocks and
$\sigma_{l,l'}$ and $\mu_l$ denote the asset covariance and average return, respectively.
An integer $w_l$ is the discrete portfolio asset position to be held,
  representing long ($w_l\ge 1$), short ($w_l\le -1$), or no-hold ($w_l=0$) position.
The parameter $\lambda$ for $0\le\lambda\le1$ adjust the asset manager's risk tolerance.
The first and second terms represent risk and return, respectively.
Thus, if $\lambda$ is large, risk is reduced.

There are two cases in the portfolio optimization problems: one in which short position is not considered,
where $w_l\in\{0, 1, \cdots, I\}$,
and the other in which it is, where $w_l\in\left\{-I/2, -I/2+1, \cdots, I/2\right\}$.
The encoding into binary for each problem is
\begin{empheq}[left={w_l=\empheqlbrace}]{alignat=2}
&  \sum_{d=1}^Df_dx_{l,d}              &\qquad& \text{without short positions}, \\
&  \frac{I}{2}-\sum_{d=1}^Df_d x_{l,d} &      & \text{with short positions}.
  \label{eq:short}
\end{empheq}
In this paper, the latter is investigated with unary encoding $f_d=1$ and $I=D$, where $D$ is defined to be even.
The cost function and constraint in Eqs. (\ref{eq:Marko}) and (\ref{eq:CpenM}) can be rewritten explicitly in the binary form as follows
\begin{eqnarray}
  E({\bm x}) &=& \frac{\lambda}{M^2} \sum_{l, l'=1}^N \sigma_{l,l'} \sum_{d,d'=1}^D\left(x_{l,d}-\frac{1}{2}\right)
  \left(x_{l',d'}-\frac{1}{2}\right)\nonumber \\
  &&+ \frac{1 - \lambda}{M} \sum_{l=1}^N \mu_l \sum_{d=1}^D\left(x_{l,d}-\frac{1}{2}\right),
        \label{eq:Markob}\\
        {\rm s.t.}&&\sum_{l=1}^N\sum_{d=1}^D x_{l,d} = M', \label{eq:Cpenb}
\end{eqnarray}
where $M'=ND/2-M$.
To convert these problems to the fermionic formulation, we can simply let $x_{l,d}\longmapsto \hat{n}_{l,d}$ for both cases in Eq. (\ref{eq:short}).

\subsection{Problem Hamiltonian with particle number conservation}
We present the problem Hamiltonian $\hat{\cal H}_p$ and
the eigenvalue problems for solving the portfolio optimization problems using quantum computation.
  The eigenvalue problems with $\hat{\cal H}_p$ are as follows
\begin{eqnarray}
  \hat{\cal H}_p\ket{\phi_{\bm x}}&=&E({\bm x})\ket{\phi_{\bm x}}\\
  {\rm s.t.\quad}  \hat{C}\ket{\phi_{\bm x}}&=&M'\ket{\phi_{\bm x}},
  \label{eq:Hconst}
\end{eqnarray}
with
\begin{eqnarray}
  \hat{\cal H}_p&=& \frac{\lambda}{M^2}\sum_{l,l'=1}^N \sigma_{l,l'}
  \sum_{d,d'=1}^D\left(\hat{n}_{l, d}-\frac{1}{2}\right)\left(\hat{n}_{l', d'}-\frac{1}{2}\right)\nonumber\\
  &&+\frac{1-\lambda}{M}\sum_{l=1}^N\mu_l\sum_{d=1}^D\left(\hat{n}_{l,d}-\frac{1}{2}\right),
  \label{eq:Hpport}\\
  \hat{C}&=&\sum_{l=1}^N\sum_{d=1}^D\hat{n}_{l,d},
  \label{eq:Nport}
\end{eqnarray}
where $E(\bm{x})$ is energy eigenvalue defined in Eq. (\ref{eq:Markob}), $M'=ND/2-M$,
and $\ket{\phi_{\bm x}}$ is the same as Eq. (\ref{eq:bit}).
The correspondence between the stock positions, variables, and quantum states for a specific stock is shown in Table \ref{tble:enc},
where the mapping to the spin-$1/2$ quantum states are also shown.
Details of mapping to spin systems are described in Sec. \ref{sec:spin}.

\begin{table}
  \caption{\label{tble:enc}
    Mapping of stock positions to variables and quantum states when short, no-hold, and long positions are considered ($D$=2).
    $\ket{\phi_l}$ ($\ket{s_{l,1}, s_{l,2}}$) is the state ket of the occupation (spin) representation for $l$-th stock, where $\ket{{\rm vac}}$ represents a vacuum.
  }
\begin{ruledtabular}          
  \begin{tabular}{ c c c  c c}
    & \multicolumn{2}{c}{Variables} &\multicolumn{2}{c}{Quantum states}\\
    \cline{2-3}\cline{4-5}
Position & $w_l$ & $x_{l,1}+x_{l,2}$ & $\ket{\phi_l}$ &$\ket{s_{l,1}s_{l,2}}_l$\\
\hline
Long & 1 & 0 & $\ket{{\rm vac}}$ & $\ket{\uparrow,\uparrow}_l$ \\\\
\multirow{2}{*}{No-hold}&\multirow{2}{*}{0} & \multirow{2}{*}{1}
& $c_{l, 2}^{\dag}\ket{{\rm vac}}$ & $\ket{\uparrow,\downarrow}_l$\\
&&& $c_{l, 1}^{\dag}\ket{{\rm vac}}$ & $\ket{\downarrow,\uparrow}_l$ \\\\
Short & $-1$ & 2 & $c_{l, 1}^{\dag}c_{l, 2}^{\dag}\ket{{\rm vac}}$ & $\ket{\downarrow,\downarrow}_l$ \\
  \end{tabular}
\end{ruledtabular}        
\end{table}

\begin{figure}[htb]
  \includegraphics[width=8.5cm]{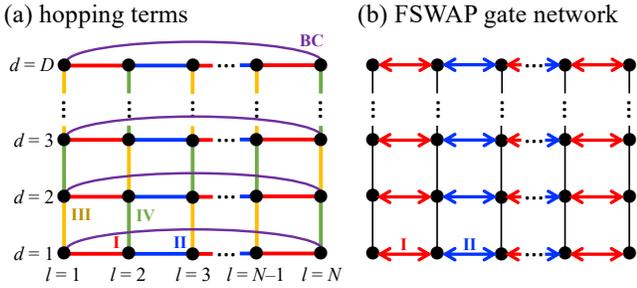}
  \caption{\label{fig:lattice}
    (a) Hopping terms of $\hat{\cal H}_t$ in Eq. (\ref{eq:Hlad}),
    and that of $\hat{U}_{\delta}(\beta)$ for $\delta=$I, II, III, IV, and BC in Eq. (\ref{eq:ulad})
    and (b) FSWAP gate network of $\hat{F}_{\delta}$ for $\delta=$I and II in Eq. (\ref{eq:F})
    on the $D$-leg ladder lattice.}
\end{figure}

\subsection{Driver Hamiltonian and initial states on $D$-leg ladder lattice}
\label{sec:Hdphi}
Following the three guidelines in Sec. \ref{sec:Hd},
we propose a driver Hamiltonian $\hat{\cal H}_d$ and an initial state $\ket{\phi_0}$ for unary encoding.
First, from condition I, $\hat{\cal H}_d$ is assumed to be a tight-binding model consisting of hopping terms.
Next, from the condition II, the hopping terms are $\hat{c}_{l,d}^{\dagger}\hat{c}_{l\pm1,d}$ and $\hat{c}_{l,d}^{\dagger}\hat{c}_{l,d\pm 1}$.
The $\hat{\cal H}_d$ designed in this way is the tight-binding model on a $D$-leg ladder lattice.
Finally, from the codition III, the \ket{\phi_0} is set to the g.s. of $\hat{\cal H}_d$ with particle number $M'$.

The tight binding Hamiltonian on the $D$-leg ladder lattice, which is adopted as $\hat{\cal H}_d$ in this paper, is defined as follows
\begin{eqnarray}
  \hat{\cal H}_{t}&=&-t^{\parallel}\sum_{l=1}^N\left(\hat{c}_{l,d}^{\dagger}\hat{c}_{l+1,d}+\hat{c}_{l+1,d}^{\dagger}\hat{c}_{l,d}\right)\nonumber\\
  &&-t^{\perp}\sum_{d=1}^{D-1}\left(\hat{c}_{l,d}^{\dagger}\hat{c}_{l,d+1}+\hat{c}_{l,d+1}^{\dagger}\hat{c}_{l,d}\right)\\
  &=&\sum_{k=1}^{N}\sum_{m=1}^D\varepsilon_{k,m}\hat{\alpha}^{\dagger}_{k,m}\hat{\alpha}_{k,m},
  \label{eq:Hlad}
\end{eqnarray}
where $t^{\parallel}$ and $t^{\perp}$ are the longitudinal and transverse hopping integrals in the tight-binding model, respectively.
The correspondence between lattice labels and hopping is shown by the colored bonds in Fig. \ref{fig:lattice} (a),
where the periodic boundary condition $\hat{c}_{N+1, d}=\hat{c}_{1,d}$ is imposed.

The energy eigenvalues $\varepsilon_{k,m}$ and the quasiparticles $\hat{\alpha}_{k,m}$ in Eq. (\ref{eq:Hlad}) have the following form
\begin{eqnarray}
  \varepsilon_{k,m}&=&-2t^{\parallel}\cos\left(\frac{2\pi}{N}k\right)-2t^{\perp}\cos\left(\frac{\pi}{D+1}m\right),
\end{eqnarray}
and
\begin{eqnarray}  
  \hat{\alpha}_{k,m}^\dagger &=& \sum_{l=1}^N\sum_{d=1}^D[\bm{\phi_0}]_{(k,m),(l,d)}\hat{c}_{l,d}^\dagger,\label{eq:alpha}\\
\ [\bm{\phi_0}]_{(k,m),(l,d)} &=& \sqrt{\frac{2}{(D+1)N}}\exp\left(i\frac{2l\pi}{N}k\right)\nonumber\\
  &&\times\sin\left(\frac{d\pi}{D+1}m\right),\label{eq:phi0unit}
\end{eqnarray}
respectively, where $k=1, 2, \cdots, N$ and $m=1, 2, \cdots, D$.
The initial state (the g.s. of $\hat{\cal H}_d=\hat{\cal H}_{t}$) in FQAOA is as follows
\begin{eqnarray}
  \ket{\phi_0} = \prod_{j=1}^{M'}\hat{\alpha}_{(k,m)_j}^{\dagger}\ket{\text{vac}},
  \label{eq:phi0lad}  
\end{eqnarray}
where the product for $j$ is taken such that $\sum_{j=1}^{M'}\varepsilon_{(k,m)_j}$ is minimal.

\subsection{Implementation on quantum circuits}\label{sec:implement}
Here we describe how to implement the ansatz shown in Eq. (\ref{eq:Ansqaoa}) on quantum circuits as Fig. {\ref{fig:QAOA}}.
The ansatz can be divided into two parts: initial state preparation $\hat{U}_{\rm init}$ and
phase rotation (mixing) unitary $\hat{U}_p(\gamma)$ [$\hat{U}_m({\beta})$].
The JW transformation \cite{Jordan} is needed to implement Fermionic operators.
In this paper, we follow the transformation shown by Babbush $et$ $al$ \cite{Babbush}.
The number of gates required for the calculation is summarized in Table \ref{tble:gateFQAOA} of Appendix \ref{sec:gate}.

\subsubsection{Initial states preperation}

The initial states $\ket{\phi_0}$ of the FQAOA on the $D$-leg ladder lattice are explicitly expressed by
the Slater determinant in Eq. (\ref{eq:phi0lad}).
The states can be prepared by quantum circuits by using Givens rotations.
Algorithms to prepare this initial state examined using quantum devices \cite{Arute2, Stanisic}.
In this paper, the algorithm of Jiang $et$ $al$. \cite{Jiang} is used to prepare the initial state on quantum circuits,
where the general Slater determinant for $n$ qubits can be computed efficiently at a circuit depth of at most $n$ for particle number $n/2$ \cite{Jiang}.
In the following, we will first present the general concepts,
and then demonstrate their implementation on actual quantum circuit through a specific example.

Following Jiang $et$ $al$. \cite{Jiang}, the general construction of $\hat{U}_{\rm init}$ is explained below.
First, to reduce the number of operations in the quantum circuit,
a unitary matrix $\bm{V}$ acts as follows:
\begin{eqnarray}
  \ket{\phi_0}&=&\prod_{j=1}^{M'}[\bm{\phi_0}\bm{\hat{c}^\dagger}]_j\ket{\rm vac}\nonumber\\
  &=&[\det(V)]^{-1}\prod_{j=1}^{M'}[{\bm V}\bm{\phi_0}\bm{\hat{c}^\dagger}]_j\ket{\rm vac},\label{eq:Vphi}
\end{eqnarray}
where $\bm{\hat{c}^\dagger} = (\hat{c}_1^\dagger, \hat{c}_2^\dagger\cdots \hat{c}_{ND}^\dagger)^T$.
The unitary matrix $\bm{V}$ is determined so that the components of the upper right triangular region of the matrix $\bm{\phi_0}$ are set to zero
and the contribution of $\det(V)$ can be ignored since it only changes the global phase.
Next, with unitary matrix ${\bm U}$ satisfying the following equation:
\begin{equation}
  [\bm{V{\phi_0}U^\dagger}]_{i,j}=e^{i\lambda_i}\delta_{i,j},\label{eq:diagVU}
\end{equation}
the following transformation is performed on Eq. (\ref{eq:Vphi})
\begin{eqnarray}
\ket{\phi_0}  &=&\prod_{j=1}^{M'}\sum_{i=1}^{ND}[\bm{V\phi_0\bm U^\dagger}]_{j,i}[\bm{U\hat{\bm c}^\dagger}]_i\ket{\rm vac}\nonumber\\
&=&e^{i\Lambda}\prod_{j=1}^{M'}[\bm{U\hat{\bm c}^\dagger}]_j\ket{\rm vac},\label{eq:Uc}
\end{eqnarray}
where the influence of $\Lambda=\sum_{j=1}^{M'}\lambda_j$ can be ignored.
The $\bm{U}$ satisfying Eq. (\ref{eq:diagVU}) can be constructed using a sequence of Givens rotations as
${\bm U}={\bm G_{N_G}\cdots{\bm G}_2{\bm G}_1}$ \cite{Jiang}, 
which can be written by 
\begin{equation}
  {\bm G_k}=
\bordermatrix{     &    &   & i & i+1 \cr
                   & 1      & \ldots & 0      & 0      &\ldots & 0 \cr
                   & \vdots & \ddots & \vdots & \vdots &\ddots & \vdots \cr  
        \quad \ \ i& 0      & \ldots & \cos\theta_k    & -e^{i\varphi_k}\sin\theta_k &\ldots & 0   \cr
               i+1 & 0      & \ldots & \sin\theta_k    &e^{i\varphi_k}\cos\theta_k     &\ldots & 0   \cr               
                   &        & \ddots & \vdots & \vdots &\ddots & \vdots \cr
                   & 0      & \ldots & 0      & 0      &\ldots & 1  \cr
            },\label{eq:matG}
\end{equation}
which is a generalization of the usual Givens rotation to a complex matrix \cite{Jiang}.
The transformation of the basis $\bm{\hat{c}^\dagger}$ by $\bm{G}_k$ can be written using the operator $\hat{\cal G}_k$ as follows:
\begin{equation}
  \hat{\cal G}_k\bm{\hat{c}^\dagger}\hat{\cal G}^\dagger_k = \bm{G}_k\bm{\hat{\bm c}^\dagger},\label{eq:Gtrans}
\end{equation}
where
\begin{equation}  
\hat{\cal G}_k= \exp\left[-i\varphi_k\left(\frac{1}{2}-\hat{n}_{i+1}\right)\right]          
          \exp\left[\theta_k \left(\hat{c}^\dagger_i\hat{c}_{i+1}-\hat{c}^\dagger_{i+1}\hat{c}_i\right)\right].\label{eq:Gop}
\end{equation}
Therefore, the transformation of the basis $\bm{\hat{c}^\dagger}$ by $\bm{U}$ in Eq. (\ref{eq:Uc}) can be written as
\begin{equation}
  \hat{\cal U}\bm{\hat{c}^\dagger}\hat{\cal U}^\dagger = \bm{U}\bm{\hat{\bm c}^\dagger},
\end{equation}
using $\hat{\cal U}=\hat{\cal G}_1\hat{\cal G}_2\cdots\hat{\cal G}_{N_G}$.
Thus, the initial state can be written in the following form
\begin{equation}
  \ket{\phi_0}=e^{-i\Phi}\hat{\cal U}\prod_{j=1}^{M'}\hat{c}^\dagger_j\ket{\rm vac},\label{eq:resphi0}
\end{equation}
where the influence of $\Phi=\sum_{k=1}^{N_G}\varphi_k/2$ can be ignored.
Finally the unitary operator used for the initial state preparation can be written as
\begin{equation}
  \hat{U}_{\rm init}=\hat{\cal G}_1\hat{\cal G}_2\cdots\hat{\cal G}_{N_G}\prod_{j=1}^{M'}\hat{c}^\dagger_j.\label{eq:opUinit}
\end{equation}

The following is an example of a circuit for initial state preparation with $M'=4$ particles on a $ND = 4\times 2$ ladder lattice.
First, as shown in Eq. (\ref{eq:Vphi}), to shorten the circuit depth, the components in the upper right triangular region of the matrix $\bm{\phi_0}$ in Eq. (\ref{eq:phi0unit})
are set to zero by the unphysical unitary transformation ${\bm V}$ as
\begin{equation}
{\bm V}\bm{\phi_0}=
\left(\begin{matrix}
  * & 7 & 4  & 2  & 1  & 0 & 0 & 0   \\
  * & * & 11 & 8  & 5  & 3      & 0 & 0   \\
  * & * & *  & 14 & 12 & 9      & 6      & 0   \\
  * & * & *  & *  & 16 & 15     & 13     & 10
\end{matrix}\right).
\label{eq:Vphi0}
\end{equation}
The matrix elements assigned numbers $j$ ($j=1,2, \cdots 16$) in the above matrix can be zeroed by acting $\bm{G^\dagger}_k$ in Eq. (\ref{eq:matG}) to the right in numeric order.
Putting a series of Givens rotations as $\bm{U}=\bm{G}_{16}\cdots \bm{G}_2\bm{G}_1$, we obtain
\begin{equation}
\bm{V\phi_0\bm U^{\dagger}} =
\left(
\begin{matrix}
   e^{i\lambda_1}  & 0  & 0     & 0   &0   & 0 & 0 & 0  \\
   0 & e^{i\lambda_2} & 0  & 0  & 0   &0   & 0 & 0  \\
   0 & 0  & e^{i\lambda_3} & 0  & 0   &0   & 0      & 0  \\
   0 & 0  & 0  & e^{i\lambda_4} & 0   &0   & 0      & 0 
\end{matrix}\right),
\end{equation}
which corresponds to Eq. (\ref{eq:diagVU}), and the initial state can be written by Eq. (\ref{eq:Uc})
  , which can be rewritten in the form of Eq. (\ref{eq:resphi0}).
Finally, the quantum circuit for preparing the initial state $\ket{\phi_0}=\hat{U}_{\rm init}\ket{\rm vac}$ can be written as
\begin{align}
  \begin{split}
    \Qcircuit @C=0.3em @R=0.7em {
\push{\rule{0.8em}{0em}}
&&&&&&&&&&&&\ket{\phi_0}\\      
&&\ustick{(l,d)}&\push{\rule{0.3em}{0em}}&&&&&&&&&\ar@{--}[]+<0.0em,0em>;[d]+<0.0em,-13em>&\\
&\lstick{\ket{0}}&\raisebox{1.2em}{1,1}\qw&\qw&\gate{X}&\qw                  &\qw                   &\qw                   &\multigate{1}{{\cal G}_7}    &\qw                  &\qw                  &\qw                  &\qw&\qw\\
&\lstick{\ket{0}}&\raisebox{1.2em}{2,1}\qw&\qw&\gate{X}&\qw                  &\qw                   &\multigate{1}{{\cal G}_{11}} &\ghost{{\cal G}_7}           &\multigate{1}{{\cal G}_{4}} &\qw                  &\qw                  &\qw&\qw\\
&\lstick{\ket{0}}&\raisebox{1.2em}{3,1}\qw&\qw&\gate{X}&\qw                  &\multigate{1}{{\cal G}_{14}} &\ghost{{\cal G}_{11}}        &\multigate{1}{{\cal G}_8}    &\ghost{{\cal G}_{4}}        &\multigate{1}{{\cal G}_{2}} &\qw                  &\qw&\qw\\
&\lstick{\ket{0}}&\raisebox{1.2em}{4,1}\qw&\qw&\gate{X}&\multigate{1}{{\cal G}_{16}} &\ghost{{\cal G}_{14}}        &\multigate{1}{{\cal G}_{12}} &\ghost{{\cal G}_8}           &\multigate{1}{{\cal G}_{5}} &\ghost{{\cal G}_{2}}        &\multigate{1}{{\cal G}_{1}} &\qw&\qw\\
&\lstick{\ket{0}}&\raisebox{1.2em}{4,2}\qw&\qw&\qw&\ghost{{\cal G}_{16}}        &\multigate{1}{{\cal G}_{15}} &\ghost{{\cal G}_{12}}        &\multigate{1}{{\cal G}_9}    &\ghost{{\cal G}_{5}}        &\multigate{1}{{\cal G}_{3}} &\ghost{{\cal G}_{1}}        &\qw&\qw\\
&\lstick{\ket{0}}&\raisebox{1.2em}{3,2}\qw&\qw&\qw&\qw                  &\ghost{{\cal G}_{15}}        &\multigate{1}{{\cal G}_{13}} &\ghost{{\cal G}_9}           &\multigate{1}{{\cal G}_{6}} &\ghost{{\cal G}_{3}}        &\qw                  &\qw&\qw\\
&\lstick{\ket{0}}&\raisebox{1.2em}{2,2}\qw&\qw&\qw&\qw                  &\qw                   &\ghost{{\cal G}_{13}}        &\multigate{1}{{\cal G}_{10}} &\ghost{{\cal G}_{6}}        &\qw                  &\qw                  &\qw&\qw\\
&\lstick{\ket{0}}&\raisebox{1.2em}{1,2}\qw&\qw&\qw&\qw                  &\qw                   &\qw                   &\ghost{{\cal G}_{10}}        &\qw                  &\qw                  &\qw                  &\qw&\qw
    }
    \end{split}
\end{align}
where
\begin{align}
  \begin{split}
    \Qcircuit @C=0.5em @R=0.7em {
&&&\lstick{i}  &\multigate{1}{{\cal G}_k} &\qw&&&
  & \lstick{i}  & \targ      & \ctrl{1}           & \targ     & \qw      & \qw \\  
&\push{\rule{0.7em}{0em}}&\push{\rule{0.5em}{0em}}&\lstick{i+1}&\ghost{{\cal G}_k}        &\qw&\push{\rule{0.9em}{0em}}&\lstick{\raisebox{1.8em}{=}}&
  \push{\rule{1.7em}{0em}}
  & \lstick{i+1}& \ctrl{-1} & \gate{R_y(-2\theta_k)} & \ctrl{-1} & \gate{R_z(\varphi_k)} & \qw &.
}
  \end{split}
\end{align}
The correspondence between $i$ and ($l,d$) is denoted in Eqs. (\ref{eq:dl}) and (\ref{eq:idl}).
Note that this operation is only valid between adjacent sites along the JW ordering in Fig. \ref{fig:JW}.

\subsubsection{Phase rotation and mixing unitary}
This section describes how to implement unitary operators of the phase rotation $\hat{U}_{p}$ and the mixing $\hat{U}_m$ on the  $D$-leg ladder lattice in Fig. \ref{fig:lattice}
  in a quantum circuit.

First, we show the implementation of $\hat{U}_p(\gamma)$.
In portfolio optimization problems, the polynominal interactions are two-body terms.
In the fermionic representation, the correspondence
$\hat{n}_{l,d}\leftrightarrow(1-\hat{Z}_{l,d})/2$
allows us to implement the interaction term as 
\begin{align}
  \label{eq:ZZInt}
  \begin{split}
    \Qcircuit @C=0.5em @R=0.7em {
&&&&\lstick{i}& \ctrl{1} & \qw & \ctrl{1} & \qw \\
&\push{\rule{7em}{0em}}&
\raisebox{2.5em}{\mbox{$\displaystyle{\exp\left[-2i\theta\left(\hat{n}_i-\frac{1}{2}\right)\otimes\left(\hat{n}_j-\frac{1}{2}\right)\right]}=$}}&
\push{\rule{8.5em}{0em}}&\lstick{j}& \targ & \gate{R_z(\theta)} & \targ & \qw&.}
  \end{split}
\end{align}

Next, we describe the implementation of
$\hat{U}_m(\beta)=\exp(-i\beta\hat{\cal H}_{t})$.
In this paper, $\hat{U}_m(\beta)$ is implemented using the Trotter decomposition
to keep the level of approximation the same as in the previous studies being compared.
The effectiveness of this method has been demonstrated in quantum devices \cite{Stanisic}.
Another technique to deal with this has been proposed using
fermionic fast fourier tansformation \cite{Verstraete, Wecker1, Wang},
which transform $\hat{\cal H}_{t}$ into a diagonal representation and performs the computation efficiently.

The followings are specific decomposition
of $\hat{U}_m(\beta)=\exp(-i\beta \hat{H}_{t})$ used in this paper,
\begin{eqnarray}
  \hat{U}_m(\beta) &=& \hat{U}_{\rm IV}(\beta)\hat{U}_{\rm III}(\beta)\hat{U}_{\rm BC}(\beta)\hat{U}_{\rm II}(\beta)\hat{U}_{\rm I}(\beta),
  \label{eq:ulad}
\end{eqnarray}
where $\hat{U}_{\delta}$ ($\delta=$ I, II, III, IV, and BC) consists of a set of commutable hopping pairs,
which are shown as color-coded bonds in Fig. \ref{fig:lattice} (a).
$\hat{U}_{\rm I, II}$ causes hopping in the leg direction along the JW ordering shown in Fig. \ref{fig:JW}.
which can be written as
\begin{eqnarray}
  \hat{U}_{\rm I} &=&
  \prod_{\substack{{d\ {\rm even}}\\{l\ {\rm even}}}}\hat{h}_{t^{\parallel}}^{i_{l,d}}(\beta)  
  \prod_{\substack{{d\ {\rm odd}}\\{l\ {\rm odd}}}}\hat{h}_{t^{\parallel}}^{i_{l,d}}(\beta),\\
  \hat{U}_{\rm II}&=&
  \prod_{\substack{{d\ {\rm even}}\\{l\ {\rm odd}}}}\hat{h}_{t^{\parallel}}^{i_{l,d}}(\beta)  
  \prod_{\substack{{d\ {\rm odd}}\\{l\ {\rm even}}}}\hat{h}_{t^{\parallel}}^{i_{l,d}}(\beta),
\end{eqnarray}
where $i_{l,d}$ is defined by Eq. (\ref{eq:idl}) and
\begin{equation}
  \hat{h}_t^i(\beta)=\exp\left[i\beta t\left(\hat{c}^{\dagger}_i\hat{c}_{i+1}+\hat{c}^{\dagger}_{i+1}\hat{c}_{i}\right)\right],\label{eq:ophop}
\end{equation}
where $t$ is the hopping integral with $t=t^{\parallel}$ or $t^{\perp}$.

The hopping terms in $\hat{U}_{\delta}(\beta)$ ($\delta=$III and IV) in Fig. \ref{fig:lattice} (a)
that are out of the JW ordering in Fig. \ref{fig:JW} can be efficiently computed using the network of fermionic swap (FSWAP) gates introduced in the literature \cite{Kivlichan}.
The implementation in this paper followed the method in Ref. \cite{Cade}.
$\hat{U}_{\rm III}(\beta)\hat{U}_{\rm IV}(\beta)$ can be written with $\hat{V}(\beta)$ to the $N$-th power as
$\hat{U}_{\rm IV}\hat{U}_{\rm III}=\hat{V}(\beta)^N$ with
\begin{eqnarray}
  \hat{V}(\beta)&=&
  \prod_{d\ {\rm even}}\hat{h}_{t^{\perp}}^{i_{1,d}}(\beta)  
  \prod_{d\ {\rm odd}}\hat{h}_{t^{\perp}}^{i_{N,d}}(\beta)
  \hat{F}_{\rm II}\hat{F}_{\rm I},\\
  \hat{F}_{\rm I}&=&
  \prod_{\substack{{d\ {\rm even}}\\{l\ {\rm even}}}}\hat{f}_{\rm swap}^{i_{l,d}}  
  \prod_{\substack{{d\ {\rm odd}}\\{l\ {\rm odd}}}}\hat{f}_{\rm swap}^{i_{l,d}},\\
  \hat{F}_{\rm II}&=&
  \prod_{\substack{{d\ {\rm even}}\\{l\ {\rm odd}}}}\hat{f}_{\rm swap}^{i_{l,d}}  
  \prod_{\substack{{d\ {\rm odd}}\\{l\ {\rm even}}}}\hat{f}_{\rm swap}^{i_{l,d}},
  \label{eq:F}
\end{eqnarray}
where $\hat{F}_{\delta}$ ($\delta =$ I and II) are depicted in Fig. \ref{fig:lattice} (b) and $\hat{f}^i_{\rm swap}$ is the FSWAP operator \cite{Wecker2, Babbush}
\begin{equation}
  \hat{f}^i_{\rm swap}=1+\hat{c}_i^\dagger\hat{c}_{i+1}+\hat{c}_{i+1}^\dagger\hat{c}_i-\hat{c}_i^\dagger\hat{c}_{i}-\hat{c}_{i+1}^\dagger\hat{c}_{i+1},
\end{equation}
which exchanges fermionic quantum states between $i$ and $i+1$ sites in the leg direction, taking into account the anti-commutation relation.

The hopping terms $\hat{U}_{\rm BC}(\beta)$ in Fig. \ref{fig:lattice} (a) can be embedded into $\hat{U}_{\rm IV}\hat{U}_{\rm III}$
where the quantum states at $l=1$ and $N$ are adjacent in the FSWAP network as
\begin{equation}
  \hat{U}_{\rm IV}\hat{U}_{\rm III}\hat{U}_{\rm BC}=\hat{V}(\beta)^{\lceil(3N-1)/4\rceil}\hat{U}_{\rm BC}\hat{V}(\beta)^{\lfloor(N+1)/4\rfloor},
\end{equation}
with
\begin{equation}
  \hat{U}_{\rm BC}(\beta)=
  \prod_{d\ {\rm even}}\hat{h}_{t^{\parallel}}^{i_{q,d}}(\beta)  
  \prod_{d\ {\rm odd}}\hat{h}_{t^{\parallel}}^{i_{p,d}}(\beta),
\end{equation}
where $p=\lfloor (N+1)/2\rfloor$ and $q=p+1$.

As an example, the first part of the $\hat{U}_m$ in Eq. (\ref{eq:ulad}) in the case of $M'=4$ particles on a $ND = 4\times 2$ ladder lattice with $t^{\parallel}=t^{\perp}\equiv t$ is shown below
\begin{align}
  \begin{split}\hspace{0.45em}
    \Qcircuit @C=0.43em @R=0.7em {
              &&&&&&&&&&&&&&\\
&\ustick{(l,d)}&&\raisebox{1.7em}{\text{$\hat{U}_{\rm I}(\beta)$}}&
\raisebox{1.7em}{\text{$\hat{U}_{\rm II}(\beta)$}}&&&\hspace{3.5em}\raisebox{1.7em}{\text{$\hat{V}(\beta)$}}&&&&\raisebox{1.7em}{\text{$\hat{U}_{\rm BC}(\beta)$}}&&&\hspace{3.5em}\raisebox{1.7em}{\text{$\hat{V}(\beta)$}}&&&\\  
&\raisebox{1.2em}{1,1}\qw&\qw&\multigate{1}{h_t(\beta)} &\qw                       &\qw&\qw                                          &\qswap    &\qw                                          &\qw                    &\qw&\qw                     &\qw&\qw                                          &\qswap    &\qw                                          &\qw                    &\qw\\
&\raisebox{1.2em}{2,1}\qw&\qw&\ghost{h_t(\beta)}        &\multigate{1}{h_t(\beta)} &\qw&\qw\hspace{0.0em}\raisebox{1.7em}{\mbox{$f$}}&\qswap\qwx&\qw                                          &\qswap                 &\qw&\multigate{1}{h_t(\beta)} &\qw&\qw\hspace{0.0em}\raisebox{1.7em}{\mbox{$f$}}&\qswap\qwx&\qw                                          &\qswap                 &\qw\\
&\raisebox{1.2em}{3,1}\qw&\qw&\multigate{1}{h_t(\beta)} &\ghost{h_t(\beta)}        &\qw&\qw                                          &\qswap    &\qw\hspace{3.4em}\raisebox{1.7em}{\mbox{$f$}}&\qswap\qwx             &\qw&\ghost{h_t(\beta)}        &\qw&\qw                                          &\qswap    &\qw\hspace{3.4em}\raisebox{1.7em}{\mbox{$f$}}&\qswap\qwx             &\qw\\
&\raisebox{1.2em}{4,1}\qw&\qw&\ghost{h_t(\beta)}        &\qw                       &\qw&\qw\hspace{0.0em}\raisebox{1.7em}{\mbox{$f$}}&\qswap\qwx&\qw                                          &\multigate{1}{h_t(\beta)}&\qw&\qw                     &\qw&\qw\hspace{0.0em}\raisebox{1.7em}{\mbox{$f$}}&\qswap\qwx&\qw                                          &\multigate{1}{h_t(\beta)}&\qw\\
&\raisebox{1.2em}{4,2}\qw&\qw&\multigate{1}{h_t(\beta)} &\qw                       &\qw&\qw                                          &\qswap    &\qw                                          &\ghost{h_t(\beta)}       &\qw&\qw                     &\qw&\qw                                          &\qswap    &\qw                                          &\ghost{h_t(\beta)}       &\qw\\
&\raisebox{1.2em}{3,2}\qw&\qw&\ghost{h_t(\beta)}        &\multigate{1}{h_t(\beta)} &\qw&\qw\hspace{0.0em}\raisebox{1.7em}{\mbox{$f$}}&\qswap\qwx&\qw                                          &\qswap                 &\qw&\multigate{1}{h_t(\beta)} &\qw&\qw\hspace{0.0em}\raisebox{1.7em}{\mbox{$f$}}&\qswap\qwx&\qw                                          &\qswap                 &\qw\\
&\raisebox{1.2em}{2,2}\qw&\qw&\multigate{1}{h_t(\beta)} &\ghost{h_t(\beta)}        &\qw&\qw                                          &\qswap    &\qw\hspace{3.4em}\raisebox{1.7em}{\mbox{$f$}}&\qswap\qwx             &\qw&\ghost{h_t(\beta)}        &\qw&\qw                                          &\qswap    &\qw\hspace{3.4em}\raisebox{1.7em}{\mbox{$f$}}&\qswap\qwx             &\qw\\
&\raisebox{1.2em}{1,2}\qw&\qw&\ghost{h_t(\beta)}        &\qw                       &\qw&\qw\hspace{0.0em}\raisebox{1.7em}{\mbox{$f$}}&\qswap\qwx&\qw                                          &\qw                    &\qw&\qw                     &\qw&\qw\hspace{0.0em}\raisebox{1.7em}{\mbox{$f$}}&\qswap\qwx&\qw                                          &\qw                    &\qw,
\gategroup{2}{6}{9}{11}{-0.8em}{^\}}\gategroup{2}{13}{9}{18}{-0.8em}{^\}}
.}
  \end{split}\label{eq:QcUm}
\end{align}
where
\begin{flalign}
\hspace{-10em}  
\Qcircuit @C=0.5em @R=.7em {
&\lstick{i}&\multigate{1}{h_t(\beta)}&\qw&\\
&\lstick{i+1}&\ghost{h_t(\beta)}&\qw&\push{\rule{3em}{0em}}}\nonumber
\end{flalign}
\begin{align}
  \begin{split}
    \Qcircuit @C=0.5em @R=.7em {
      &&&\lstick{i}& \gate{R_x({-\pi/2})} &\ctrl{1}& \gate{R_x(-\beta t)} &\ctrl{1} & \gate{R_x({\pi/2})}  & \qw \\
      &\raisebox{2.0em}{=}&\push{\rule{2.5em}{0em}}&\lstick{i+1}& \gate{R_x({\pi/2})}  &\targ   & \gate{R_z(\beta t)}  &
\targ    & \gate{R_x({-\pi/2})} &\qw&,
}
\end{split}\label{eq:Qchop}
\end{align}
and
\begin{align}
  \begin{split}
    \Qcircuit @C=0.5em @R=1.0em {
&&&\lstick{i}  &\qw&\qswap &\qw&&&
  & \lstick{i}  & \qswap      & \qw      & \ctrl{1} & \qw      & \qw \\  
      &&\push{\rule{5.0em}{0em}}&\lstick{i+1}
      &\qw\hspace{0.2em}\raisebox{1.9em}{\mbox{$f$}}&\qswap\qwx  &\qw&\push{\rule{1.0em}{0em}}
      &\hspace{-1.5em}{\raisebox{1.9em}{\mbox{$=$}}}&
  \push{\rule{2.0em}{0em}}
  & \lstick{i+1}& \qswap \qwx & \gate{H} & \targ    & \gate{H} & \qw &.
}
  \end{split}\label{eq:Qcfswap}
\end{align}
The correspondence between $i$ and ($l,d$) is denoted in Eqs. (\ref{eq:dl}) and (\ref{eq:idl}).
Note that Eqs. (\ref{eq:Qchop}) and (\ref{eq:Qcfswap}) are only valid between adjacent sites along the JW ordering in Fig. \ref{fig:lattice} (b).

\subsection{Numerical evaluation of FQAOA}
In this section, the validity of our proposed computational framework FQAOA is investigated through numerical simulations.
First, we examine the proposed driver Hamiltonian in the conventional adiabatic time-evolution framework.
Next, by comparing the statistical data obtained by FQAOA with other methods, we evaluate the performance of FQAOA assuming the use of NISQ devices.

\subsubsection{Computational frameworks to be compared}
\label{sec:spin}
Our proposed FQAOA is compared with two computational frameworks: the conventional $X$-QAOA \cite{Farhi1} with soft constraint and
  the $XY$-QAOA with hard constraint \cite{Hodson}.
  These are based on spin representations with
  spin-$1/2$ operator $\hat{s}^\alpha_i$ for directions $\alpha=x,y$, and $z$ at the $i$ site.
    In the $X$-QAOA, the initial state $\ket{\phi_0}$, which does not satisfy the constraints,
    is transferred to the approximate optimal solution indicated by the problem Hamiltonian $\hat{\cal H}_p$ with a penalty term $\hat{\cal H}_{\rm pen}$ added.
    In the $XY$-QAOA, on the other hand, the $\hat{\cal H}_{\rm pen}$ is not required,
    because an eigenstate of the constraint operator $\hat{C}$ is chosen as an $\ket{\phi_0}$
    and a mixer $\hat{U}_m^{XY}=\exp(-i\hat{\cal H}_d)$
    satisfying the commutation relation $[\hat{\cal H}_d, \hat{C}]=0$ in Eq. (\ref{eq:condI}) is applied.
  
The problem Hamiltonian $\hat{\cal H}_p$ and the constraint operator $\hat{C}$ are obtained by
applying the mapping $\hat{n}_{l,d}\leftrightarrow 1/2-\hat{s}_{l,d}^z$ to Eq. (\ref{eq:Hpport}) and (\ref{eq:Nport}) as
\begin{eqnarray}
	\hat{\cal H}_p &=& \frac{\lambda}{M^2} \sum_{l, l'=1}^N \sigma_{l,l'}\sum_{d, d'=1}^D\hat{s}^z_{l,d}\hat{s}^z_{l',d'}
        - \frac{ 1 - \lambda}{M} \sum_{l=1}^N \mu_l \sum_{d=1}^D\hat{s}^z_{l,d},\label{eq:Hpsp}\nonumber\\
        \label{eq:Hpsp}\\
        \hat{C}&=&\frac{ND}{2}-\sum_{l=1}^N\sum_{d=1}^D \hat{s}^z_{l,d},\label{eq:Nsp}
\end{eqnarray}
respectively.
The driver Hamiltonian $\hat{\cal H}_d$ and initial state $\ket{\phi_0}$ explained below are summarized in Table \ref{tble:compare}.

In the $X$-QAOA, $\hat{\cal H}_d=\hat{\cal H}_X$ is used, where
\begin{equation}
  \hat{\cal H}_X= -2\sum_{l=1}^{N}\sum_{d=1}^{D}\hat{s}^x_{l,d}.\label{eq:HX}
\end{equation}
The g.s. of $\hat{\cal H}_X$ is simply expressed as $\{[\ket{\uparrow}+\ket{\downarrow}]/\sqrt{2}\}^{\otimes ND}$,
which is used as the initial state $\ket{\phi_0}$.
Since $[\hat{\cal H}_d, \hat{C}]\ne 0$ 
a penalty Hamiltonian $\hat{\cal H}_{\rm pen}$ has to be added to the $\hat{\cal H}_p$ as
\begin{eqnarray}
  \hat{\cal H}'_p&=&\hat{\cal H}_p+A\hat{\cal H}_{\rm pen},\label{eq:Hpprime}\\
 \hat{\cal H}_{\rm pen} &=& \left(\sum_{l=1}^N\sum_{d=1}^D s_{l,d}^z-M \right)^2,\label{eq:Hpenspin}  
\end{eqnarray}
where $A$ is a parameter that adjusts the strength of the penalty.

\begin{figure}[htb]
  \includegraphics[width=5cm]{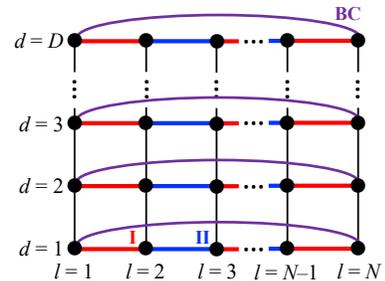}
  \caption{\label{fig:exchange}
Exchange terms of $\hat{U}^{XY}_{\delta}(\beta)$ for $\delta$=I, II, and BC on a $D$ circle lattice.
  }
\end{figure}

Another comparison (referred as $XY$-QAOA-I and II) is based on the quantum alternating operator ansatz \cite{Hadfield1},
which Hodson $et$ $al$. applied to the portfolio optimization problem \cite{Hodson}.
We here introduce the following $XY$ Hamiltonian on a $D$ circle lattice shown in Fig. \ref{fig:exchange} as
\begin{equation}
   \hat{\cal H}_{XY}= -2\sum_{l=1}^{N}\sum_{d=1}^{D}\left(\hat{s}^x_{l,d}\hat{s}^x_{l+1,d}+\hat{s}^y_{l,d}\hat{s}^y_{l+1,d}\right),
  \label{eq:HXY}
\end{equation}
where the condition $[\hat{\cal H}_{XY}, \hat{C}]=0$ is satisfied.
In this case, the hard constraint can be imposed on the ansatz in Eq (\ref{eq:Ansqaoa}) by preparing an initial state that satisfies the constraints.
In Ref. \cite{Hodson}, the special case of $D=2$ is applied.
For implementation of the mixer $\hat{U}^{XY}_m(\beta)=\exp(-i\beta\hat{\cal H}_{XY})$,
the following Trotter decomposition is applied
\begin{eqnarray}
  \hat{U}_m^{XY}(\beta) &=& \hat{U}^{XY}_{\rm BC}(\beta)\hat{U}^{XY}_{\rm II}(\beta)\hat{U}^{XY}_{\rm I}(\beta),  \label{eq:UXY}
\end{eqnarray}
where $U_{\delta}^{XY}$ ($\delta $= I, II, and BC) consists
of sets of commutable exchange pairs, which are shown as color-coded bonds in left panel of Fig. \ref{fig:exchange}.

Next, we explain the initial state $\ket{\phi_0}$ used in the methods $XY$-QAOA-I and II.
Hodson $et$ $al$. \cite{Hodson} adopted $\ket{\phi_0}$ with long positions for $M$ specific stocks and 
no-hold positions for the others as
\begin{equation}
  \ket{\phi_{\rm I}}=\bigotimes_{l=1}^{M}  \ket{\uparrow,\uparrow}_l
  \bigotimes_{l=M+1}^N\left[\frac{1}{\sqrt{2}}(\ket{\uparrow,\downarrow}_l+\ket{\downarrow,\uparrow}_l)\right],
  \label{eq:phiI}
\end{equation}
where the correspondence between spin states and positions is shown in Table \ref{tble:enc}.
We refer to their method using this initial state as $XY$-QAOA-I,
in which the simulation results strongly depend on the initial stock holdings.
Therefore, we introduce another method $XY$-QAOA-II,
  in which the initial stock position of $XY$-QAOA-I are superposed in all cases as
\begin{eqnarray}
  \ket{\phi_{\rm II}}&=&\frac{1}{\sqrt{N!}}\sum_{\bm P}\bigotimes_{l=1}^{M}  \ket{\uparrow,\uparrow}_{P_l}\nonumber\\
  &&\bigotimes_{l=M+1}^N\left[\frac{1}{\sqrt{2}}(\ket{\uparrow,\downarrow}_{P_l}+\ket{\downarrow,\uparrow}_{P_l})\right],
  \label{eq:phiII}
\end{eqnarray}
where $\bm{P}$ is permutation of $N$ symbols for the index $l$.
Note that since neither $\ket{\phi_{\rm I}}$ nor $\ket{\phi_{\rm II}}$ is a g.s. of the $\hat{\cal H}_{XY}$,
the QAA calculation cannot be applied.

\begin{table}
  \caption{\label{tble:compare}
Comparison of calculation methods,
where the driver Hamiltonian $\hat{\cal H}_d$, initial states $\ket{\phi_0}$, and optimization methods of variational parameters are shown.
In FQAOA, Broyden-Fletcher-Goldfarb-Shanno (BFGS) and conjugate gradient (CG) method give the same results.
  }
\begin{ruledtabular}          
  \begin{tabular}{ c c c c c }
Method & $\hat{\cal H}_d$ & $\ket{\phi_0}$ & \shortstack{Variational \\ optimizer} \\
\hline
$X$-QAOA      & $\hat{\cal H}_X$    & g.s. of $\hat{\cal H}_X$ & stochastic BFGS \\\\
$XY$-QAOA-I \cite{Hodson}  & \multirow{2}{*}{$\hat{\cal H}_{XY}$} & $\ket{\phi_{\rm I}}$   & Nelder-Mead\\
$XY$-QAOA-II  &  & $\ket{\phi_{\rm II}}$
& stochastic BFGS\\\\
FQAOA & $ \hat{\cal H}_{t}$ & g.s. of $\hat{\cal H}_{t}$ & BFGS or CG \\
	\end{tabular}
	\label{tab:init}
        \end{ruledtabular}        
\end{table}

\subsubsection{Computational details}
In the following, we compare our results with a previous study by Hodson $et$ $al$ \cite{Hodson}.
To compare with the previous study, the case where short positions can be taken ($D=2$) for eight stocks ($N=8$) and
the total number of stocks held ($M=4$) are used in the calculations.
In Eq. (\ref{eq:Marko}), the parameters $\sigma_{l,l'}$ and $\mu_{l}$ are the values in Fig. 2 and Table IV of Ref. \cite{Hodson}, respectively.
We also set $\lambda=0.9$ and $A=0.003$ according to the paper.
The energy is scaled by $W$, which is the energy range of $\hat{\cal H}_p$ in Eq. (\ref{eq:Hpport}) under the constraints of Eq. (\ref{eq:Hconst}).
The hopping parameters in $\hat{\cal H}_{t}$ are set to $t^{\parallel}=t^{\perp}=W/W_t$,
where the $W_t$ is the energy range of $\hat{\cal H}_t$ at $t^{\parallel}=t^{\perp}=1$.

Numerical simulations have been performed using the fast quantum simulator qulacs \cite{qulacs}.
For the parameter optimization to obtain $E_p({\bm \gamma}^*,{\bm \beta}^*)$ in Eq. (\ref{eq:optEp}),
the Broyden-Fletcher-Goldfarb-Shanno (BFGS) and conjugate gradient (CG) algorithm has been applied in our FQAOA.
On the other hand, in $X$-QAOA and $XY$-QAOA-II,
the BFGS with 100 basin hopping is applied to avoid trapping in the local minimum.
The Nelder-Mead method has been applied in the previous studies ($XY$-QAOA-I) \cite{Hodson}.
The methods used for the parameter optimization are summarized in the fourth column of the Table \ref{tab:init}.

\subsubsection{Simulation results using fixed-angle FQAOA}

\begin{figure}[t]
  \includegraphics[width=8.5cm]{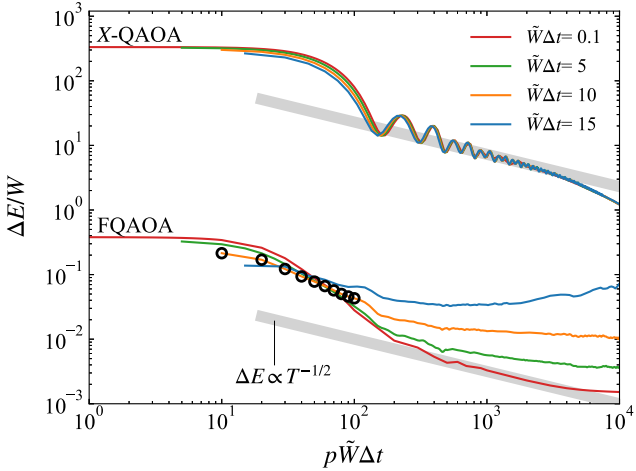}
  \caption{\label{fig:QAA}
    Comparison of the residual energy $\Delta E=E_p({\bm \gamma}^{(0)}, {\bm \beta}^{(0)})-E_{\rm min}$
    for fixed-angle FQAOA and $X$-QAOA,
    where (${\bm \gamma}^{(0)}, {\bm \beta}^{(0)}$)
    is obtained by the discrete-time QAA in Eq. (\ref{eq:gamma0beta0}) and
    $E_{\rm min}$ is the ground state energy of $\hat{\cal H}_p$ under the constraint.
    As a reference, the data
      for $p=1, 2, \dots, 10$ at $\tilde{W}\Delta t=10$ are shown by the open circles.
    Two bold gray lines represent the power law of $\Delta E\propto T^{-1/2}$ for execution time $\tilde{W}T=p\tilde{W}\Delta t$.
  }
\end{figure}

Hereafter we will show the simulation results of our proposed FQAOA.
First, we compare the results of the fixed-angle FQAOA with that of the $X$-QAOA,
where the variational parameters are fixed to
  $({\bm \gamma},{\bm \beta})=({\bm \gamma}^{(0)},{\bm \beta}^{(0)})$
  in Eq. (\ref{eq:gamma0beta0}) for various $\tilde{W}\Delta t$.
  We then show that FQAOA is reduced to QAA in the large $p$ limit.

The simulation results of the residual energy $\Delta E=E_p({\bm \gamma}^{(0)}, {\bm \beta}^{(0)})-E_{\rm min}$
for various $p$ by using fixed-angle QAOA are shown in Fig. \ref{fig:QAA}.
  This corresponds to a time-discretized QAA with execution time $T=p\Delta t$
  and the discretized time width $\Delta t$ is adjusted for verification.
  These $T$ and $\Delta t$ are scaled by $\tilde{W}$, which is the energy range of $\hat{\cal H}'_p$ in Eq. (\ref{eq:Hpprime}) for $X$-QAOA
  and $\tilde{W}=W$ for FQAOA. 

According to the results in Fig. \ref{fig:QAA}, the computational error of fixed-angle FQAOA is smaller than
that of the conventional method $X$-QAOA for all $p\tilde{W}\Delta t$ region.
This is because the FQAOA ansats strictly satisfies the constraint at all approximation levels $p$,
  whereas that of $X$-QAOA treats the constraints approximately
  and the penalty imposed on states out of the solution space increases the energy.
  The data in FQAOA for sufficiently small $\tilde{W}\Delta t=0.1$ lie on the power law $\Delta E(T)\propto T^{-1/2}$,
  which confirms the asymptotic approach to the g.s. by QAA \cite{Suzuki, Keck, Kudo}.
Also in $X$-QAOA, the asymptotic lines are parallel and their pre-factors are 1,000 orders of magnitude larger.
Therefore, we can conclude that the FQAOA method,
  which intrinsically treats the constraints as particle number conservation laws,
  is superior to the conventional $X$-QAOA method.  
  As a point of reference, open circles represent the results of fixed-angle FQAOA for $p=1, 2, \dots ,10$ at $\tilde{W}\Delta t =10$.
  The results obtained here at $p=1$ are superior to the previous study of $XY$-QAOA-II as will be seen next.

\begin{figure*}[htb]
  \includegraphics[width=17cm]{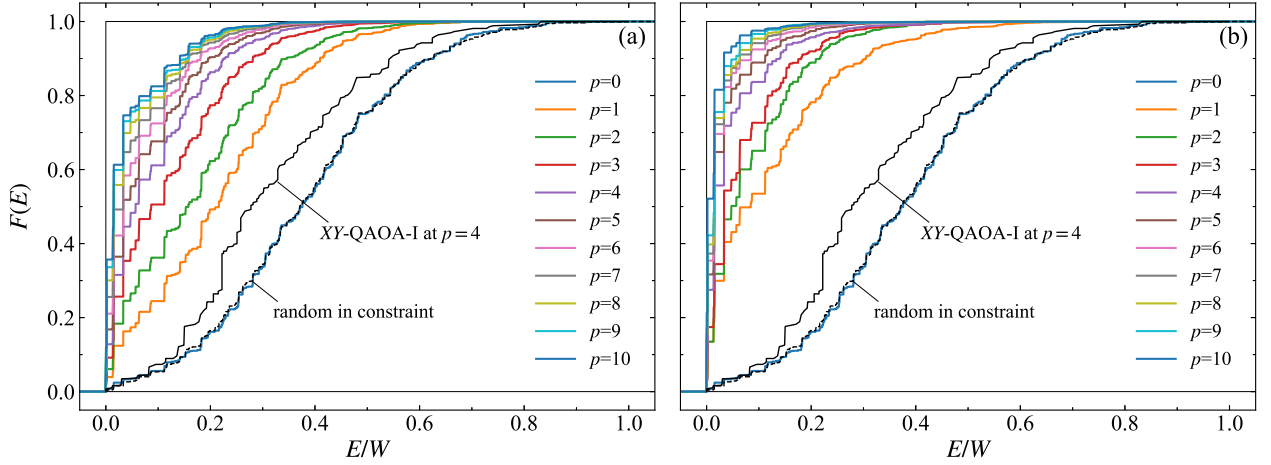}
\caption{
  \label{fig:CumP}
  Energy dependence of cumulative probability $F(E)$ in Eq. (\ref{eq:FE}) obtained by using
  (a) fixed-angle QAOA with $({\bm \gamma}, {\bm \beta})=({\bm \gamma}^{(0)}, {\bm \beta}^{(0)})$ in Eq. (\ref{eq:gamma0beta0}) at $W\Delta t =10$ and (b) FQAOA with $({\bm \gamma}, {\bm \beta})=({\bm \gamma}^*, {\bm \beta}^*)$,
    which is obtained from $({\bm \gamma}^{(0)}, {\bm \beta}^{(0)})$ via optimization.
The data shown as $XY$-QAOA-I are extracted from Fig. 8 in Ref. \cite{Hodson} and
`random in constraint' indicates the distribution of random sampling under the constraints.
  Rectangles represent the cumulative probability in the exact ground state.
}
\end{figure*}

\begin{figure}[htb]
  \includegraphics[width=8.5cm]{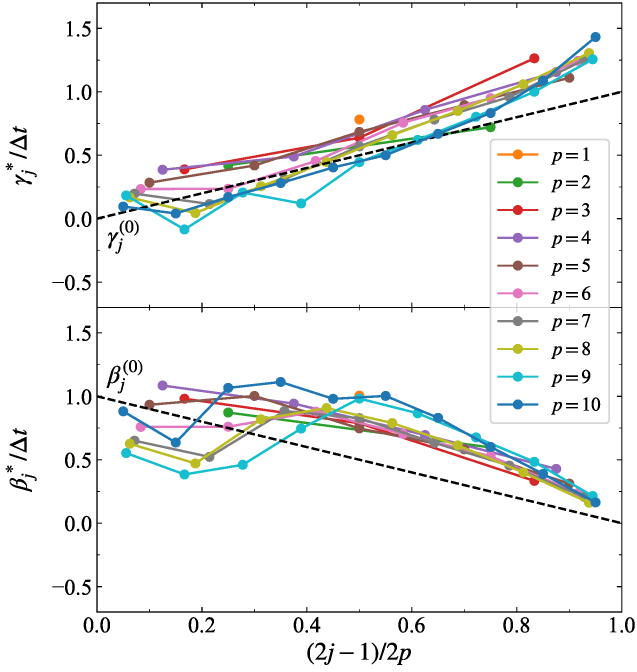}
\caption{
  \label{fig:gammabeta}
  The variational parameters (${\bm \gamma}^*, {\bm \beta}^*$),
  which are obtained from $({\bm \gamma}^{(0)}, {\bm \beta}^{(0)})$ via optimization.
  The optimized parameters (${\bm \gamma}^*, {\bm \beta}^*$) (solid lines) for FQAOA obtained by Broyden-Fletcher-Goldfarb-Shanno (BFGS)
  or conjugate gradient (CS) algorithm.
  The initial parameters (${\bm \gamma}^{(0)}, {\bm \beta}^{(0)}$) in Eq. (\ref{eq:gamma0beta0}) at $W\Delta t =10$
    are shown by dashed lines.
}
\end{figure}

\begin{figure*}[htb]
  \includegraphics[width=17cm]{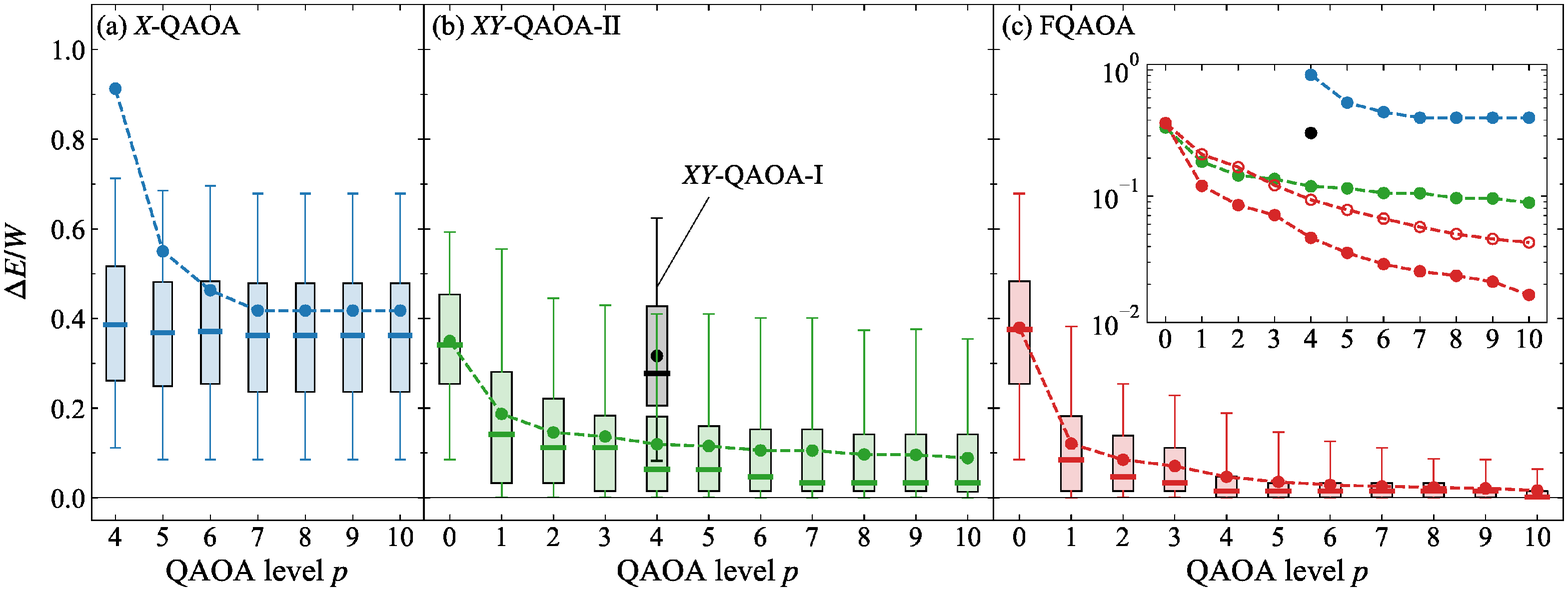}
  \caption{\label{fig:box}
    Statistical analysis of computational errors $\Delta E$ in Eq. (\ref{eq:sampling}) depending on the QAOA level $p$.
    In each box plot, the box shows quartiles, the divider is the median and
    the ends of the whisker indicate a range where the cumulative probability is between 0.05 to 0.95.
    Circles with broken lines represent the expectation values $\Delta E/W = [E_p({\bm \gamma}^*, {\bm \beta}^*)-E_{\rm min}]/W$,
    where $E_{\rm min}$ is the g.s. energy of $\hat{\cal H}_p$ under the constraint.
    The inset in (c) shows expectation values $\Delta E/W$ in log scale,
    where $\Delta E$ using $E_p({\bm \gamma}^{(0)}, {\bm \beta}^{(0)})$ at $W\Delta t=10$ are shown by the open circles.
}
\end{figure*}

\subsubsection{Simulation results using FQAOA with paremeter optimization}
  Here we present simulation results from the framework of FQAOA.
  First, to see the effect of parameter optimization,
  the results of the cumulative probability $F(E)$ obtained by
  (a) fixed-angle FQAOA and (b) FQAOA with parameter optimization are shown in Fig. \ref{fig:CumP}.
  The scaled variational parameters used in each calculation are presented in Fig. \ref{fig:gammabeta}.

The $F(E)$ is expressed by the following equations
\begin{eqnarray}
  F(E)&=& \int_0^E P(E') dE',\label{eq:FE}
\end{eqnarray}
with probability distribution $P(E)=\partial F(E)/\partial E$, which is derived from
\begin{eqnarray}
  P(E)&=& \sum_{\bm{x}}
  P_{\bm{x}}({\bm \gamma},{\bm \beta})
  \delta\left\{E-\left[E(\bm{x})-E_{\rm min}\right]\right\},
\end{eqnarray}
where $P_{\bm x}({\bm \gamma},{\bm \beta})$ is the probability of observing bit string ${\bm x}$ defined by Eq. (\ref{eq:Px}).
By a simple calculation, the error $\Delta E$ can be obtained by the following equation
\begin{equation}
  \Delta E= \int_0^{W} dE'\left[1- F(E')\right].\label{eq:DEFE}
\end{equation}
Since the area enclosed by the rectangle and the resulting curve in Fig. \ref{fig:CumP} corresponds to the $\Delta E$,
the smaller this area is, the smaller the expected value of the error.

First we investigate the results of fixed-angle FQAOA shown in Fig. \ref{fig:CumP} (a).
The $F(E)$ at $p=0$ is derived from the initial state $\ket{\phi_0}$, where the probability distribution is close to that of the randomly sampled under the constraint,
so the $\ket{\phi_0}$ is represented by a superposition of all classical states satisfying the constraint.
  The $F(E)$ of fixed-angle FQAOA at $p = 1$ is larger than the value of the previous study $XY$-QAOA-I at $p = 4$ shown in Fig. 8 of the Ref. \cite{Hodson},
  so the computational error $\Delta E$ in the former is smaller than that in the latter.
  As a result, the fixed angle FQAOA at $p = 1$ outperforms the $XY$-QAOA-I at $p = 4$ \cite{Hodson} in the following three respects:
  the computational accuracy, the number of gate operations, and no need for variational parameter optimization.
  The actual number of gate operations for fixed angle FQAOA at $p=1$ is about half the number of those for $XY$-QAOA-I at $p=4$;
  specifically, the number of operations for single- and two-qubits is 556 and 656 for the former, respectively, while the latter is 936 and 1092.
These values are derived from the Table \ref{tble:gate} in Appendix \ref{sec:gate}.

  We now turn to the results in (b).
  We see that the area between the rectangle and the curves decreases at each $p$ compared to that in (a),
    indicating that the
  $\Delta E$ in Eq. (\ref{eq:DEFE}) decreases for all $p$.
  In particular, $P(E)=\partial F(E)/\partial E$ increases in the low energy region.
  This means that the parameter optimization of FQAOA increases the probability of realization of low-energy (low-cost) states.
  Indeed, it can be seen that the probability of obtaining a solution in the low-energy region
    $0\le E\le W/100$, represented by $F(E/100)$,
    increases from $0.357$ to $0.521$ by parameter optimization at $p=10$.

We would like to make mention of the parameter optimization in Eq. (\ref{eq:optEp}).
In the previous study of $XY$-QAOA-I \cite{Hodson}, the Nelder-Mead method was applied,
which is not realistic to implement in real devices because it requires multiple trials for initial parameter settings.
In addition, there is no guarantee that the accuracy will improve with an increase in $p$ for a small number of trials \cite{Hodson}.
On the other hand, in the FQAOA of this study, by setting $({\bm \gamma}^{(0)},{\bm \beta}^{(0)})$ as initial parameters,
the accuracy of the fixed-angle FQAOA is at least guaranteed. 
This appropriate initial parameter settings also allow the parameter to be updated efficiently in a single local search using the BFGS and CG methods.
Figure \ref{fig:gammabeta} shows the results of the updated variational parameters.
The optimized parameters scale well for increasing $p$,
suggesting that $({\bm \gamma}^{(0)},{\bm \beta}^{(0)})$ is a good starting point for any $p$.
The value of $E({\bm \gamma}^*,{\bm \beta}^*)$ is checked to be in agreement with the value obtained by the stochastic method with 100 times basin hopping.

Finally, we compare the calculation accuracy of FQAOA with other calculation methods $X$-QAOA and $XY$-QAOA-I and II.
In Fig. \ref{fig:box}, the statistical data of error $\Delta E$ are shown by box-and-whisker diagram,
which are obtained from the following sufficiently large number of sampling
\begin{equation}
  \Delta E=\sum_{\substack{{\bm x}\\{\rm sampling}}} [E({\bm x})-E_{\rm min}]P_{\bm x}({\bm \gamma^*},{\bm \beta^*}),\label{eq:sampling}
\end{equation}
in a noiseless environment,
where $P_{\bm x}({\bm \gamma}^*,{\bm \beta}^*)$ is the probability in Eq. (\ref{eq:Px})
with the optimized variational parameter.
The $\Delta E$ in Fig. \ref{fig:box} decreases monotonically for each method as $p$ increases.
In particular, the $\Delta E$ of FQAOA shown in Fig. \ref{fig:box} (c) can be significantly suppressed compared to the other methods.
In the case of $X$-QAOA, the energy increase due to the penalty term in the problem Hamiltonian $\hat{\cal H}'_p$ in Eq. (\ref{eq:Hpprime}) causes $\Delta E> W$ for $p\le3$.
For $p\ge$7, it becomes difficult to find the optimal (${\bm \gamma}^*, {\bm \beta}^*$) by parameter optimization.
Next we look at the $XY$-QAOA-I and II results in Fig. \ref{fig:box} (b) with black and orange plots, respectively.
The former $\Delta E$ takes a value smaller than $X$-QAOA,
however, it is strongly dependent on the initial state of the particular set of stocks holdings.
To eliminate the initial state dependence, the latter uses an initial state that is a symmetric operation on the former one shown in Eq. (\ref{eq:phiII}).
In this case, the computational error is reduced, however the convergence for increasing $p$ is worse than for FQAOA.
Compared to these results, for $p \ge 1$, the error of FQAOA is the smallest and its variance is also largely suppressed.

\begin{table}
  \caption{\label{tble:res}
Comparison of computational accuracy,
where the error $\Delta E/W = [E_p({\bm \gamma}^*,{\bm \beta}^*)-E_{\rm min}]/W$
and the probability of realization $F(W/100)$ of low-energy states for $0\le E\le W/100$ are shown.
In the fixed-angle FQAOA, $({\bm \gamma}^*,{\bm \beta}^*)$ is replaced by $({\bm \gamma}^{(0)},{\bm \beta}^{(0)})$.
  }
\begin{ruledtabular}          
  \begin{tabular}{ c c c c c }
    & \multicolumn{2}{c}{$\Delta E/W$} & \multicolumn{2}{c}{$F(W/100)$} \\
    \cline{2-3}\cline{4-5}    
    method&$p=4$&$p=10$&$p=4$&$p=10$\\
    \hline
$X$-QAOA                  & 0.913 & 0.418 & 0.004 & 0.005\\\\
$XY$-QAOA-I \cite{Hodson} & 0.3   & ---   & 0.007 & ---\\
$XY$-QAOA-II              & 0.120 & 0.089 & 0.092 & 0.237 \\\\
fixed-angle FQAOA         & 0.094 & 0.043 & 0.127 & 0.357 \\
FQAOA                     & 0.047 & 0.017 & 0.275 & 0.521  \\
	\end{tabular}
        \end{ruledtabular}
\end{table}

  A summary of comparison is shown in the inset of Fig. {\ref{fig:box}} (c),
  where the fixed-angle FQAOA results are also shown by open circles.
  It can be seen that the fixed-angle results of FQAOA for $p\ge 3$
  are below the computational errors of the parameter-optimized $XY$-QAOA-I and II.
  The parameter-optimized FQAOA results have the smallest error compared to all other methods for a given $p$ in the region of $p\ge 1$.
  Quantitative comparisons at $p=4$ and $10$ are shown in Table \ref{tble:res}.
  The error $\Delta E$ and the probability of realization $F(W/100)$ of the low-energy state monotonically decrease and increase, respectively,
  from the $X$-QAOA to the FQAOA, indicating that the calculation accuracy improves in terms of both $\Delta E$ and $F(W/100)$.
  Comparing the results of FQAOA at $p=4$ with those of previous study $XY$-QAOA-I \cite{Hodson},
  the calculation error is reduced to 1/6 and the probability of realization of low-energy states is increased by a factor of 40.
  The largest low-energy probability is $F(W/100)=0.521$ for the parameter-optimized FQAOA at $p=10$.

  The FQAOA also shows superiority in variational parameter optimization calculations.
  In the $XY$-QAOA methods, the fixed-angle QAOA reduced to QAA is not applicable.
  Therefore, stochastic methods have to be applied for parameter optimization due to the lack of a policy for setting initial parameters,
  which increases the computation time by the number of initial parameter sampling.
On the other hand, as we have seen above, FQAOA provides good accuracy without parameter optimization,
and moreover, the parameter-optimized results as variational algorithm can be obtained without initial parameter sampling.
This property is a significant advantage in performing calculations on real devices.
\\

\section{Summary and Discussion}

In this paper, we proposed fermionic QAOA (FQAOA) for constrained optimization problems.
This algorithm obtains the optimized solution by changing the physical system from the non-local fermions described by the driver Hamiltonian to the localized fermions corresponding to the optimization problem, where the constraint condition is regarded as the number of fermions.
In this paper, we express the design guideline of the driver Hamiltonian in a fermionic formulation and propose a strategy to set the ground state of the driver Hamiltonian as the initial state.
We also proposed to use the parameters of the fixed-angle QAOA corresponding to the time-discretized QAA as the initial variational parameters.
As a demonstration of this algorithm, a portfolio optimization problem was taken up.
The fixed-angle QAOA calculation confirms that the residual energy calculated by the FQAOA and the $X$-QAOA ansatz decays at similar powers, however, the FQAOA method, which essentially treats the constraint as a particle number conservation law, is superior to the conventional $X$-QAOA method.
The resulting computational accuracy of the fixed-angle FQAOA at $p=1$ exceeded
that of the parameter-optimized $XY$-QAOA at $p=4$ in the previous study \cite{Hodson} in about half the number of gate operations.
Furthermore, the parameter-optimized FQAOA at $p=4$ improved the probability of achieving low-energy states
by a factor of 40 compared to that in the $XY$-QAOA at $p=4$ \cite{Hodson}.

In the present paper, non-interacting fermions were set as the initial state.
On the other hand, the initial state preparation using Givens rotation allows the implementation of any Slater determinant.
Thus, for example, the Hartree-Fock solution can be set as the initial state.
  In the most promising field of quantum chemical calculations, such calculations have been realized on quantum devices \cite{Arute2}.
  Therefore, our algorithm is expected to significantly improve the computational accuracy of constrained combinatorial optimization problems by using the quantum chemical computing tools such as OpenFermion \cite{openfermion}, 
  both for NISQs and FTQCs.

We note here that the unary encoding was applied in the portfolio optimization problem.
  For other encodings, FQAOA can be performed by preparing the initial state in a way that satisfies Eq. (\ref{eq:Cpenb}).
  However, since $f_d\ne 1$, the constraint $\sum_{l}x_{l,d}=m_d$ is imposed independently for each $d$,
  where $\sum_d m_d=M'$.
Correspondingly, since there are multiple initial states that depend on the assignment of $m_d$, multiple FQAOAs must be attempted to search for the optimal solution.
In addition, one-hot encoding and other efficient encodings may impose restrictions on embedding general variables into binary values.
  In this case, FQAOA can be performed in the same way by imposing constraints in encoding.\\

Finally, it is necessary to mention the quantum advantage.
This is not limited to QAOA, but is an open problem in the field of quantum computing.
In the context of QAOA, for example, warm-start techniques have been shown to improve the performance of the algorithm for some optimization problems \cite{Egger, Tate}, where the quantum algorithm can inherit the performance guarantees of classical algorithms.
However, the effectiveness of this strategy is a future challenge as it depends on the available resources.

\begin{acknowledgments}
  T.Y. thank H. Kuramoto and Y. Takamiya for valuable discussions.
K.F. is supported by MEXT Quantum Leap Flagship Program (MEXTQLEAP) Grants No. JPMXS0118067394 and No. JPMXS0120319794. This work is supported by JST COI-NEXT program Grant No. JPMJPF2014.
\end{acknowledgments}

\appendix

\section{Quantum Adiabatic Algorithm}
\label{sec:qaa}
We explain the general QAA \cite{Kadowaki, Farhi1, Aharonov}. It consists of a problem Hamiltonian $\hat{\cal H}_p$ and a driver Hamiltonian $\hat{\cal H}_d$ and satisfies $[\hat{\cal H}_p, \hat{\cal H}_d]\ne 0$.
Taking the initial state $\ket{\phi_0}$ as the g.s. of $\hat{\cal H}_d$,
the approximate g.s. $\ket{\psi(T)}$ of $\hat{\cal H}_p$ for a sufficiently large $T$ is obtained by the following calculation:
\begin{eqnarray}
 \ket{\psi(T)}= \hat{U}(T)\ket{\phi_0},
  \label{eq:UTphi}
\end{eqnarray}
where $\hat{U}(T)$ is the time evolution operator for execution time $T$.
Suppose the schedule of the time-dependent Hamiltonian is $\hat{\cal H}(t)=(1-t/T)\hat{\cal H}_d+(t/T)\hat{\cal H}_p$,
then the time-evolution operator $\hat{U}(t)$ can be written in the following form
\begin{eqnarray}
  \hat{U}(t) &=& T_{t'}\exp\left\{ -i\int_0^t \left[\left(1-\frac{t'}{t}\right)\hat{\cal H}_d+\frac{t'}{t}\hat{\cal H}_p\right] dt'\right\},\nonumber\\
    \label{eq:UT}
\end{eqnarray}
where $T_t'$ is the time ordering product for $t'$.

According to Ref. \cite{Farhi2}, we introduce an approximate form of Eq. (\ref{eq:UT}),
which is obtained by discretizing time $t$ by $\Delta t$ and then performing the following Trotter decomposition:
\begin{eqnarray}
\hat{U}(T) &\sim& 
\prod_{j=1}^{p}\exp\left[-i\left(1-\frac{t_j}{T}\right)\hat{\cal H}_d\Delta t\right]\exp\left[ -i\frac{t_j}{T}\hat{\cal H}_p\Delta t\right],\nonumber\\
\label{eq:Trot}
\end{eqnarray}
where $t_j = (2j-1)\Delta t/2$ with $T=p\Delta t$.
The approximate g.s. and the expectation value of $\hat{\cal H}_p$ are given by
\begin{equation}
  \ket{\psi(T)}=\hat{U}(T)\ket{\phi_0}\sim\prod_{j=1}^{p}\hat{U}_m(\beta^{(0)}_j)\hat{U}_p(\gamma^{(0)}_j)\ket{\phi_0},
\end{equation}
and
\begin{equation}
  E(T)=\bra{\psi(T)}{\cal H}_p\ket{\psi(T)}\sim E_p({\bm \gamma}^{(0)},{\bm \beta}^{(0)}),
  \label{eq:Eqaa}
\end{equation}  
with
\begin{eqnarray}
  \gamma_j^{(0)} &=& \frac{2j-1}{2p}\Delta t, \\  
  \beta_j^{(0)}  &=& \left(1-\frac{2j-1}{2p}\right)\Delta t.
\end{eqnarray}
The accuracy of the approximate time evolution operator in Eq. (\ref{eq:Trot}) is characterized by a execution time $T=p\Delta t$ and a discrete time width $\Delta t$.
By taking the $\Delta t$ small enough and the $T$ large enough, the circuit inevitably becomes deeper and
the accuracy of the approximation can be improved.
The expectation value $E(T)$ in Eq. (\ref{eq:Eqaa}) can be used to evaluate the performance of ansatz.
Using a well-designed ansatz,
a pure quantum algorithm, such as the present fixed-angle FQAOA in Fig. \ref{fig:CumP} (a) and the previous studies \cite{Cao}, gives good performance.

  \section{Comparison of Gate Counts}\label{sec:gate}

\begin{table}[htb]
  \caption{\label{tble:gateFQAOA}
Numbers of single- and two-qubit gates of $\hat{U}_{\rm init}$, $\hat{U}_p$, and $\hat{U}_m$ in FQAOA ansatz.}
\begin{ruledtabular}          
  \begin{tabular}{ c c c }
    Operator & Single-qubit gate& Two-qubit gate\\
    \hline
    $\hat{U}_{\rm init}$&
$\displaystyle{\frac{1}{4}(ND+2M+2)(ND-2M)}$     
    &
$\displaystyle{\frac{3}{4}\left[(ND)^2-4M^2\right]}$\\\\    
$\hat{U}_{p}$  & $\displaystyle{\frac{1}{2}ND(ND+1)}$
    & $ND(ND-1)$    \\\\
$\hat{U}_{m}$ & $2N^2D+10ND-6N$ 
    & $2N^2D+2ND-2N$  
	\end{tabular}
        \end{ruledtabular}
\end{table}

The number of gate operations for FQAOA obtained by counting the number of gates in section \ref{sec:implement} is shown in Table \ref{tble:gateFQAOA}.
  As a comparison, the counts for the four methods ($X$-QAOA, $XY$-QAOA-I and II, and FQAOA)
  at $D=2$ are also shown in Table \ref{tble:gate}.
For $\hat{U}_p=\exp(-i\gamma\hat{\cal H}_p)$, all methods are implemented with the same quantum circuit, and the number of single- and two-qubit gate operations are as follows

\begin{table}
  \caption{\label{tble:gate}
Comparison of the gate counts of $\hat{U}_{\rm init}$ and $\hat{U}_m$ at $D = 2$.
The numbers of single- and two-qubit gates for each method are shown in the upper and lower rows, respectively.}
\begin{ruledtabular}          
  \begin{tabular}{ c c c c }
    Method& $\hat{U}_{\rm init}$& $\hat{U}_{m}$\\
    \hline
    $X$-QAOA              & $2N$ & $2N$  \\
                          & $0$ & $0$  \\\\
$XY$-QAOA-I \cite{Hodson} & $2(N-M)$&$12N$\\
                          & $N-M$&  $4N$\\\\
$XY$-QAOA-II              & $4NM-4M^2+4N+1$ &$12N$\\
                          & $5M(N-M)$      &  $4N$\\\\
FQAOA                     & $(N+M+1)(N-M)$   & $4N^2+14N$   \\
                          & $3(N^2-M^2)$ & $4N^2+2N$   \\    
	\end{tabular}
        \end{ruledtabular}
\end{table}

\[N(2N+1), \qquad 2N(2N-1),\]
respectively.
The only difference between $XY$-QAOA I and II is the initial state.
In $XY$-QAOA-I, $N-M$ triplet states are generated for specific unowned shares as shown in Eq. (\ref{eq:phiI}).
On the other hand, $XY$-QAOA-II requires the generation of a superposition of unowned shares as shown in Eq. (\ref{eq:phiII}).
This initial state can be prepared by the following unitary operation
\begin{eqnarray}
  \hat{U}_{\rm init} &=&  \exp{\left(i\frac{\pi}{2}\sum_{l=1}^N\hat{s}^z_{l,1}\right)}\nonumber\\
  &&\times\exp{\left[-i\frac{\pi}{2}\sum_{l=1}^N\left(\hat{s}^x_{l,1}\hat{s}^x_{l,2}+\hat{s}^y_{l,1}\hat{s}^y_{l,2}\right)\right]}\nonumber\\
  &&\times\hat{\cal C}_{N,N-M},\label{eq:UinitXY}
\end{eqnarray}
where $\hat{\cal C}_{N,N-M}$ is the operator that generates a Dicke state of hamming weight $N-M$ on $N$ qubits with $d=2$.
The number of gate operations for single- and two-qubits in $\hat{\cal C}_{N,N-M}$, respectively are as follows \cite{Mukherjee}
\[4NM-4M^2-2N+1, \qquad 5NM-5M^2-2N.
\]
Also, since $\sum_{l=1}^N\hat{s}^z_{l,1}$ is commutative with both $\hat{\cal H}_{p}$ and $\hat{\cal H}_{XY}$, the $R_z$ gate in the first line of Eq. (\ref{eq:UinitXY}) can be ignored.

\end{document}